# Consonant gemination in Italian: the nasal and liquid case

Maria Gabriella Di Benedetto (*), Luca De Nardis (**)

(*) Radcliffe Institute for Advanced Study at Harvard University, Cambridge, USA, and Sapienza University of Rome, Rome, Italy

(**) DIET Department, Sapienza University of Rome
Rome, Italy

{mariagabriella.dibenedetto, luca.denardis}@uniroma1.it

**Abstract**
All Italian consonants affected by gemination, that is affricates, fricatives, liquids, nasals, and stops, were analyzed within a project named GEMMA that lasted over a span of about 25 years. Results of the analysis on stops, as published in (Esposito, A., and Di Benedetto, M. G. (1999). "Acoustic and Perceptual Study of Gemination in Italian Stops," The Journal of the Acoustical Society of America, ASA, Vol. 30, pp. 175-185) showed that the main acoustic cue to gemination in Italian was closure duration, while frequency and energy domain parameters were not significantly affected by gemination. This paper - the first of a set of two covering all remaining consonants - addresses nasals and liquids; its companion paper addresses affricates and fricatives. Results on nasals and liquids confirm the findings on stops, in particular that the primary acoustic cue to gemination in Italian is durational in nature and corresponds to a lengthened consonant duration. Results also show an inverse correlation between consonant and pre-consonant vowel durations which is, however, also present when considering singleton vs. geminate word sets separately, indicating a sort of duration compensation between these segments to eventually preserve rhythmical structures; this inverse correlation is reinforced when considering singleton and geminate sets combined. Classification tests of singleton vs. geminate consonants show that, for both nasals and liquids, best classification scores are obtained when consonant duration is used as a classification parameter. Although slightly less performing, the ratio between consonant and pre-consonant vowel durations is also a potential good candidate for automatic classification of geminate vs singleton nasals and liquids in Italian.

## 1. Introduction

Gemination can be defined as the clustering of a single consonant into a 'double' or geminate consonant. This phenomenon plays a major role in Italian, a language in which gemination is contrastive and therefore several words change their meaning due to the presence or absence of gemination of one consonant in the word. Words belonging to minimal pairs are orthographically distinguished by a double grapheme of the geminate consonant (for example: papa (pope) vs. pappa (baby food), or pala (shovel) vs. palla (ball)). Native Italian speakers have a natural attitude in producing disyllabic words of minimal pairs identified by the presence or absence of consonant gemination. In Italian, moreover, gemination can be also observed across neighbouring words of the same intonational phrase, giving rise to a phenomenon that is peculiar of the Italian language, called "raddoppiamento sintattico."

The identification of acoustic correlates of gemination in Italian, and the verification of their perceptual relevance, is a longstanding research challenge. Previous studies addressed Italian stops (Rossetti, 1993; Rossetti, 1994; Esposito and Di Benedetto, 1999), based on the analysis of speech materials consisting in VCV vs. VCCV words. Results showed that consonant closure and pre-consonant vowel durations were affected by gemination. In particular, when gemination was present the pre-consonant vowel duration was decreased while consonant closure duration was increased, suggesting that speakers may tend to preserve the rhythmic structure of the word. Similar observations were also reported by (Rochet and Rochet, 1995) and (Pickett *et al.*, 1999), where the latter also observed some kind of constancy in the phenomenon across speaking rates. Evidence for relational acoustic relevance was also found in Japanese (Hirata and Whiton, 2005).

The role of durational cues in relation to gemination was highlighted in a seminal paper by Fujisaki et al. (1973). Gemination was investigated in several other languages; evidence for consonant duration as the main acoustic cue to gemination was also found in stops and fricatives in Lebanese (Al-Tamimi and Khattab, 2011; Khattab and Al-Tamimi, 2014; Al-Tamimi and Khattab, 2015), in Hindi (Shrotriya *et al.*, 1995), in Cypriot Greek (Arvaniti, 1999; Arvaniti and Tserdanelis, 2000; Tserdanelis and Arvaniti, 2000), in Persian stops (Hansen, 2004),  in three languages of Indonesia (Cohn *et al.*, 1999), in Swedish and Iraqi Arabic (Hassan, 2003), in Japanese stops (Idemaru and Guion, 2008) and in Berber (Louali and Maddieson, 1999; Ridouane, 2007),  although in Berber geminate stops lack their singleton counterparts. Other acoustic cues to gemination such as shortening of the pre-consonant vowel in the presence of gemination were found in Italian (Esposito and Di Benedetto, 1999) and (Turco and Braun, 2016). Although vowel shortening was reported to be less evident in running speech, in particular for fast speaking rates, (Pickett *et al.*, 1999) observed a systematic difference in vowel duration also in fast speaking rates, at least for stressed vowels. Shortening of the preconsonantal vowel in non-isolated words was also reported in (Turco and Braun, 2016). Speaking rate also affected consonant lengthening (Pickett *et al*, 1999), causing a marked reduction of the difference between geminated and singleton consonants as the rate increases. Pre-consonant vowel shortening was also observed in Berber (Ridouane, 2007), but neither in Persian (Hansen, 2004) nor in Arabic (Hassan, 2003).

Acoustic cues related to the frequency domain - rather than durational in nature - were also investigated in Italian, albeit significant variations in singletons vs. geminates were only observed for specific consonants. In particular, a significant decrease of F1 in the liquid consonant /l/, along with an increase in F2 and F3 in pre-stressed and unstressed positions, in presence of gemination, were observed in (Payne, 2005). An impact of gemination on frequency domain parameters was also observed in an Austronesian language, Pattani Malay (Abramson, 1998). The study of Pattani Malay focused on the analysis of fundamental frequency (F0) variations with gemination of word-initial consonants. Findings were that F0 varied with gemination, although not for all consonantal classes. In particular, F0 in nasal consonants was not affected by gemination, while the opposite was true for stops, as also confirmed in a perceptual experiment (Abramson, 1999). A Dravidian language, Malayalam (Local and Simpson, 1999), stands somewhat apart from others, for spectral and temporal properties seem to be equally relevant in characterizing gemination.

The speech group at Sapienza University of Rome, Italy, has been active in tackling the problem of finding acoustic cues to gemination in the Italian language for many years; the Gemination project GEMMA (Di Benedetto, 2000; GEMMA, 2019) started at Sapienza in 1992, with the ambition of analyzing gemination for all Italian consonants occurring in both singleton and geminate forms. The analyzed consonants were stops, liquids, fricatives, nasals, and affricates. The first extensive publication output of the GEMMA project addressed gemination in stop consonants (Esposito and Di Benedetto, 1999).

Beyond stops, all published materials appeared either in abstracts or in currently out-of-print journals; results for liquids were presented in a meeting of the Acoustical Society of America (Argiolas et al., 1995), while for all other consonants in the former and no longer available (since mid-2011) copyright-free web journal named European Student  Journal of Language and Speech "WEB-SLS" (fricatives: Giovanardi and Di Benedetto, 1998; nasals: Mattei and Di Benedetto, 2000; affricates: Faluschi and Di Benedetto 2000).

The purpose of this manuscript, and of its companion paper (Di Benedetto and De Nardis, 2020), is to extend, revisit, and contextualize, the work originally presented in the above papers. Novel contributions of the present submissions include the analysis on liquid consonants, exhaustive statistical analyses on time, frequency and energy domain parameters, and the analysis of time domain parameters as potential test variables for the classification of singleton vs. geminate words. As such, this paper and its companion paper provide a comprehensive assessment and offer to the speech research community the benefit of accessing to results and data that do not appear anymore in a published form.

The reference value of the material is reinforced by providing, as integral part of this revisit, full access to the entire database on which the study has been founded. This database is a unique case of Italian consonants in VCV vs. VCCV words. A detailed description of the database is provided in Section 2, along with details on speech material for nasals and liquids, analyzed in the present paper. Acoustic analyses and statistical tests are presented in Section 3. Results of acoustic analysis are reported in Section 4. Section 5 provides a discussion and comparison of results for nasals vs. liquids, as well as the results of classification tests for singleton vs. geminate words. Section 6 draws conclusions and highlights future directions of research.

## 2. Speech materials
### 2.1 The GEMMA database

The speech materials analyzed in the present work, and in its companion paper (Di Benedetto and De Nardis, 2020), are part of the GEMMA project database (GEMMA, 2019). This database is composed of disyllabic words, i.e. vowel–consonant–vowel (VCV) in the nongeminate case and vowel–consonant–consonant–vowel (VCCV) in the geminate case. The consonants in the words are stops (/b/, /d/, /g/, /p/, /t/, /k/), affricates (/tʃ/, /dʒ/, /ts/, /dz/), fricatives (/f/, /v/, /s/), nasals (/m/, /n/) and liquids (/l/, /r/), that is all consonants of the Italian language that are generally accepted as appearing in both single or geminate forms in intervocalic position. The case of affricates is, however, a debatable one, as will be further discussed in the companion paper. The vowels in the words are /a, i, u/, that is a subset of Italian vowels /a, e, ɛ, i, o, ɔ, u/. Words are symmetric with respect to vowel.

Six adult Italian native speakers raised and living in Rome (Italy), three men and three women aged from twenty-four to fifty, participated in the recordings. The speakers were pronunciation defectless and free of evident dialectal inflexions. As suggested in previous studies (Payne, 2006), the Roman accent, although quite distinctive, is phonologically very close to Standard Italian. It is by the way interesting to note that Mairano and De Iacovo (2019) postulate a progressive standardization of the Italian language based on a comprehensive study of the impact of regional variations of Italian on gemination.

The words in the GEMMA database were pronounced in isolation and not in carrier sentences, in order to limit the effect of factors such as intonation (Rossetti, 1993, 1994), and in particular to mitigate variations in intonation between different speakers. Although it can in fact be expected that a same speaker will repeat a same sentence with similar intonation, this may not be the case across speakers. This difference is expected to be much less evident in isolated words, especially in relatively short VCV/VCCV words.

Words were written on cards that were presented to the speaker by the operator. Cards were shuffled after each recording session. No distractors were included in the recording protocol, leading to the possibility for speakers to guess the aim of the experiment, and thus involuntarily introducing a bias in the experiment. This risk was mitigated by the supervision of the recording sessions by an acoustically trained person, also in charge of pointing out evident mispronunciations and prompting a new recording when needed. Furthermore, the use of multiple repetitions helped attenuating the impact of residual biases in the recorded material.

The speech materials of the GEMMA database were recorded in the Speech Laboratory of the INFOCOM Department (now DIET Department) at the University of Rome 'La Sapienza' (Italy) using professional equipment, in a sound-treated room. The entire set of words was recorded three times in three different recording sessions, leading to three repetitions for each word and for each speaker. In case of evident mispronunciations, the speaker was compelled to repeat the word.

The distance of the speakers from the microphone was monitored during the recording sessions and was kept at about 20 cm. Speakers were asked to maintain their natural speaking style in order to mitigate the impact of variations in emission levels and tempo. An analysis of the average standard deviation of utterance duration variations within each speaker across repetitions revealed that this is contained within about 3-5% of the average utterance duration. As expected, variations across speakers were larger, with an average standard deviation of about 10% of the average utterance duration, and were mitigated by including six different speakers.

The words were then digitized using the UNICE software produced by VECSYS (Vecsys, 2019). Speech signals were filtered at 5 kHz, sampled at 10 kHz, and each sample was quantized with 16 bits. Each signal was then stored by UNICE as a .sig file containing the samples and a companion .key file with information on sampling rate and quantization.

The GEMMA database is now available under an open source Creative Commons license; the original UNICE file doublets describing each speech signal were converted into .wav files using the sox open source utility, in order to offer a wide access to the material (GEMMA, 2019). The top folder of the database contains a README file providing a detailed description of its organization, briefly summarized as follows. The database is organized in five folders, one for each family of consonants: folder "Affricates" for affricates, folder "Fricatives" for fricatives, folder "Liquids" for liquids, folder "Nasals" for nasals and folder "Stops" for stops. Each of the above folders is further organized into six folders, one for each speaker, named "FS1", "FS2", "FS3", for the three female speakers, and "MS1", "MS2", "MS3" for the three male speakers. Acronyms for the six speakers are stored in the README file. Each speaker folder contains the files for the three repetitions for that specific consonant set; the generic file name is in the form "<Word><Repetition><Speaker>.wav," e.g., the first repetition for the word "iffi" for the first female speaker is named "IFFI1FS1.wav".

## 2.1. Nasals and liquids speech materials

In the Italian language, the set of nasal consonants that is generally accepted as appearing in both singleton and geminate forms is /m, n/, since /ɲ/ appears only in the geminate form, while liquids that appear in the Italian language in both singleton and geminate forms are /l, r/ (Muljacic, 1972). Table I shows the set of words in the database containing nasal and liquid consonants, where consonants in the geminated form are represented by a double grapheme of the consonant. Given the number of speakers (6 speakers), the number of repetitions (3 repetitions), the number of symmetrical vowel contexts (3 vowel contexts), the number of consonants (2 nasals, 2 liquids) and the forms (singleton vs. geminate), a total of 6x3x3x2x2=216 words were recorded for liquid consonants, and 216 for nasal consonants.

|   | Nasals | | | | Liquids | | | |
|---|---|---|---|---|---|---|---|---|
|   | **m** | | **n** | | **l** | | **r** | |
| **a** | ama | amma | ana | anna | ala | alla | ara | arra |
| **i** | imi | immi | ini | inni | ili | illi | iri | irri |
| **u** | umu | ummu | unu | unnu | ulu | ullu | uru | urru |

**Table I** - Set of words of the GEMMA database that contain nasal and liquid consonants. Singleton consonants are indicated by /m, n/ and /l, r/, while geminate consonants are indicated by /mm/, /nn/, /ll/, /rr/.

## 3. Measurements and statistical tests

Software tools used to perform measurements and statistical tests are described in Section 3.1.1. Measurements of parameters were taken at specific times and frames that are defined in Section 3.1.2. Time domain parameters are described in Section 3.1.3. Frequency domain and energy domain parameters are described in Sections 3.1.4 and 3.1.5, respectively. Finally, Section 3.1.6 describes the statistical tests that have been adopted to analyse the statistical significance of parameters.

### 3.1.1. Software tools

The speech analysis was carried out using version 3.1 of the software xkl, developed by Dennis Klatt, for Linux and macOS environments (Klatt, 1984).
Statistical analyses were carried out using IBM SPSS Statistics version 25 (IBM Corp., 2017) for ANOVA tests and the Statistics and Machine Learning toolboxes of MATLAB R2019b (MATLAB, 2019) for correlation and classification tests described in Section 3.1.6, both in a macOS environment.

### 3.1.2. Reference times and reference frames

The analyzed parameters were measured at specific instants in time, called reference times, that correspond to relevant acoustic events within the word. The identification of reference times was made based on the specific characteristics of each consonant. Reference times were determined by visual inspection of waveforms and spectrograms and can be listed as follows (see Fig. 1):

- Vowel 1 onset time ($V1_{onset}$) – The pre-consonant vowel onset time, $V1_{onset}$, was identified by the appearance of a glottal pulse followed by other regular glottal pulses.
- Vowel 1 offset time ($V1_{offset}$) – The pre-consonant vowel offset time, $V1_{offset}$, was identified as the time at which higher formants disappear.
- Vowel 2 onset time ($V2_{onset}$) – The post-consonant vowel onset time, $V2_{onset}$, was identified as the time instant at which a glottal pulse appeared, and an abrupt shift in formants was visible. The decision was also supported in specific cases by a short-term energy analysis and in a few cases by direct listening.
- Vowel 2 offset time ($V2_{offset}$) – The post-consonant vowel offset time, $V2_{offset}$, was typically matched with the disappearance of the second and higher formants. In specific cases, mostly in words including the [u] vowel, $V2_{offset}$ was set as the time at which signal amplitude decreased below about 90% of its peak value.
- Consonant onset time ($C_{onset}$) – coinciding with $V1_{offset}$.
- Consonant offset ($C_{offset}$) – coninciding with $V2_{onset}$.

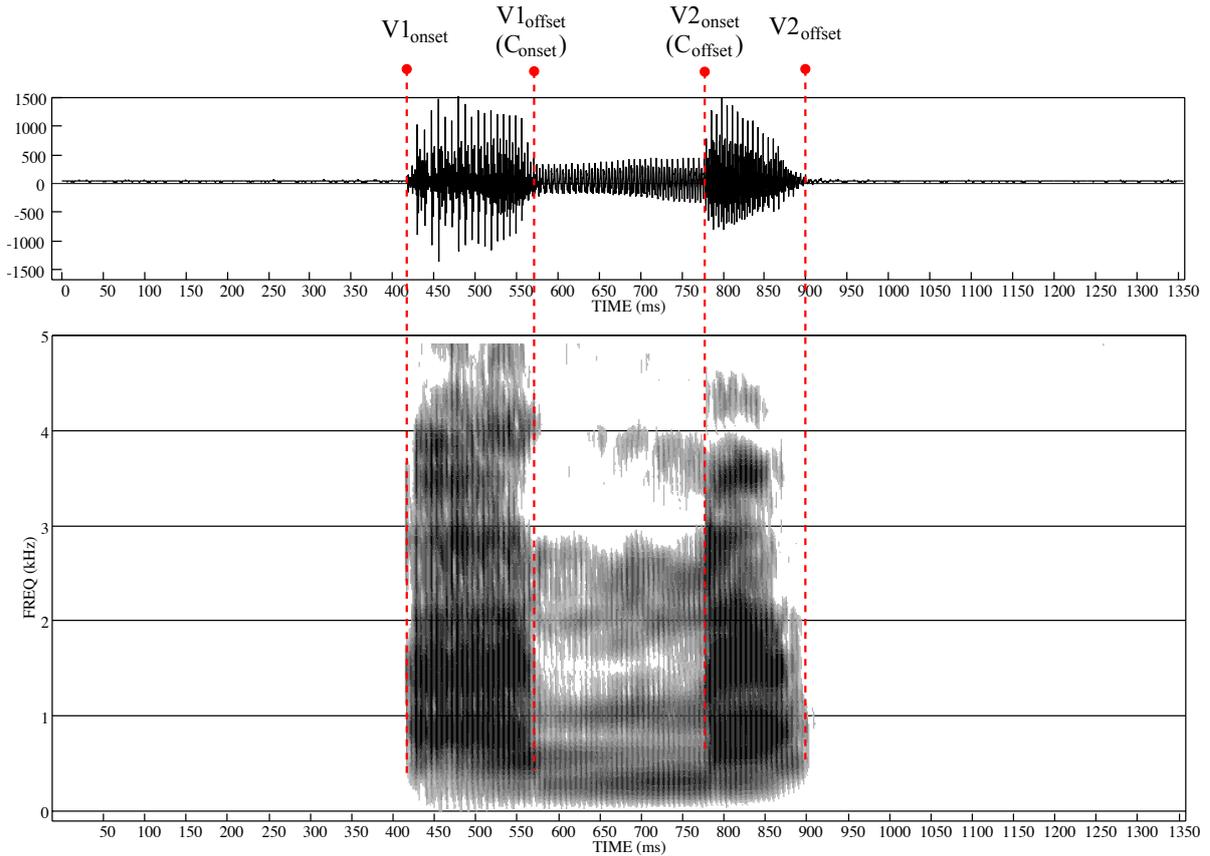

**Figure 1 -** Reference times for the computation of the acoustic parameters. $V1_{onset}$: reference time corresponding to onset of pre-consonant vowel; $V1_{offset}$: offset of pre-consonant vowel, corresponding to consonant onset $C_{onset}$; $V2_{onset}$: onset of post-consonant vowel, corresponding to the consonant offset $C_{offset}$; $V2_{offset}$: offset of post-consonant vowel.

A set of reference frames, each consisting of 256 samples, was also defined, with respect to reference times. Figure 2 shows the reference frames, that are defined as follows:

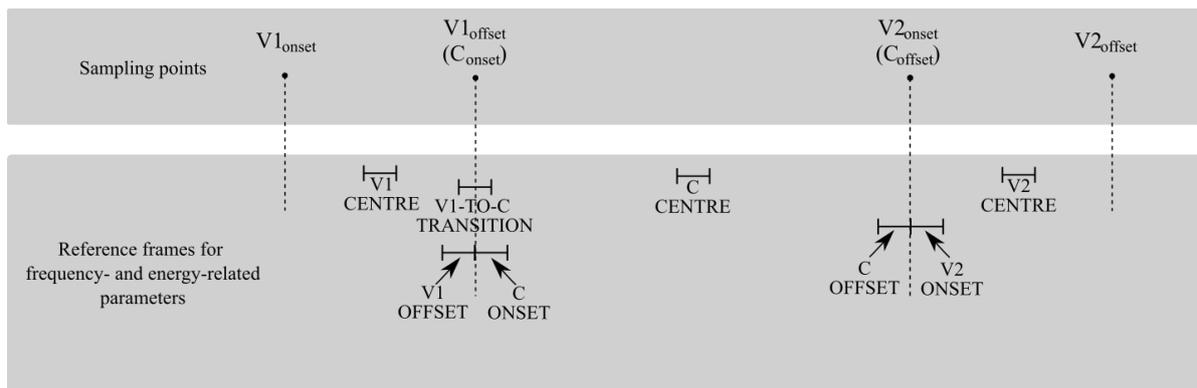

**Figure 2** – Reference frames defined with respect to the reference times of Figure 1, each containing 256 samples.

- V1 CENTRE – frame located at V1 center, i.e. centered on $\frac{V1_{onset}+V1_{offset}}{2}$;
- V1 OFFSET – frame located at the offset of V1, right before $V1_{offset}$;
- V1-TO-C TRANSITION – frame located at the transition between V1 and C, centered on $V1_{offset}$;
- C ONSET – frame located at the onset of the consonant, i.e. starting at $V1_{offset}$;
- C CENTRE– frame located at C center, i.e. centered on $\frac{V1_{offset}+C_{offset}}{2}$;
- C OFFSET – frame located at the offset of the consonant, i.e. ending at $C_{offset}$;
- V2 ONSET – frame located at the onset of V1, i.e. starting at $V2_{onset}$;

- V2 CENTRE – frame located at the center of V2, i.e. centered on $\frac{V2_{onset}+V2_{offset}}{2}$.

### 3.1.3. Time domain parameters
Figure 3 shows the time domain parameters, defined as follows:
- duration of the pre-consonant vowel V1d=V1$_{offset}$-V1$_{onset}$;
- duration of the consonant Cd=C$_{offset}$-C$_{onset}$;
- duration of the post-consonant vowel V2d=V2$_{offset}$-V2$_{onset}$;
- duration of the entire word Utd=V2$_{offset}$-V1$_{onset}$.

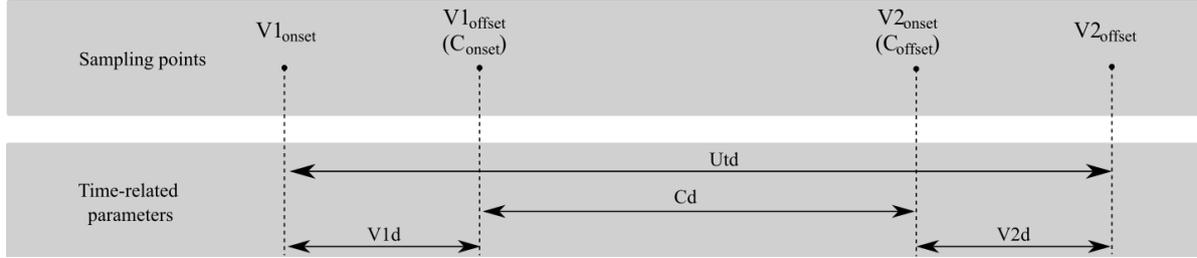

**Figure 3** – Time domain parameters defined with respect to reference times of Figure 1. V1d: duration of first vowel; Cd: duration of consonant; V2d: duration of second vowel; Utd: duration of the entire word.

### 3.1.4. Frequency domain parameters
An exhaustive analysis of the impact of gemination on frequency parameters was also carried out. In order to perform the analysis in the frequency domain, speech signals were pre-emphasized with a pre-emphasizing filter with α=0.95 and windowed using a Hamming window of 256 samples. Spectrograms, DFT (Discrete Fourier Transform) and LPC (Linear Predictive Coding) spectra were examined and compared to extract the following parameters:
- Fundamental frequency F0;
- First three formant frequencies F1, F2 and F3.

The above parameters were evaluated with respect to the reference frames as follows (see Figure 2 for reference):
- V1 CENTRE: F0, F1, F2 and F3;
- V1 OFFSET: F0, F1, F2 and F3;
- V1-TO-C TRANSITION: F0, F1, F2 and F3;
- C ONSET: F0 (both nasals and liquids), F1, F2 and F3 (liquids only);
- C CENTRE: F0 (both nasals and liquids), F1, F2 and F3 (liquids only);
- C OFFSET: F0 (both nasals and liquids), F1, F2 and F3 (liquids only);
- V2 ONSET: F0, F1, F2 and F3;
- V2 CENTRE: F0, F1, F2 and F3.

### 3.1.5. Energy domain parameters
The following energy domain parameters were defined:
- total energy of V1, $E_{totV1}$, defined as $E_{totV1}=\sum|X_i|^2$, where $X_i$ is i-th sample falling in the time interval [V1$_{onset}$, V1$_{offset}$], corresponding to the duration of V1;
- average power of V1, $P_{V1} = E_{totV1}/N_{V1}$, where $N_{V1}$ is the number of samples over [V1$_{onset}$, V1$_{offset}$];
- total energy of C, $E_{totC}$, computed as for V1, but over C duration [C$_{onset}$, C$_{offset}$];
- average power of C, indicated as $P_C$, and computed from $E_{totC}$ as for $P_{V1}$, but dividing by the number of samples within the interval [C$_{onset}$, C$_{offset}$];
- instantaneous energy at V1 CENTRE, $E_{iV1cent}$, defined as $E_{iV1cent}=\sum|X_i|^2$, where $X_i$ is i-th sample belonging to the V1 CENTRE reference frame;
- instantaneous energy at the transition V1-to-C, $E_{iV1-C}$, computed as $E_{iV1}$ but in the V1-TO-C TRANSITION reference frame;
- instantaneous energy at C CENTRE, $E_{iCcent}$, and instantaneous energy at C OFFSET, $E_{iCoff}$, computed as $E_{iV1cent}$.

All energy domain parameters listed above were expressed in logarithmic form ($10\log_{10}(x)$).

### 3.1.6. Statistical tests
The following statistical tests were performed (Dillon W.R. and Goldstein M., 1984):
- Repeated measurements ANOVA and multi-factor univariate ANOVA, to determine whether average values of parameters presented statistically significant differences between different groups of words;
- Spearman Rank Correlation Coefficient, used to detect correlations between the different parameters;
- Pearson's Correlation coefficient, used to determine correlation when the relation between parameters is linear. The proximity of Spearman and Pearson's coefficients indicates that the relation between parameters is both monotonic and linear;
- Maximum Likelihood Classification (MLC) test, to determine parameters to classify singleton vs. geminate words.

## 4. Results

### 4.1. Results on nasals
#### 4.1.1. Results in the time domain
Figure 4 shows the values of V1d, Cd, V2d and Utd averaged over repetitions and speakers for nasal consonants [m, n], and corresponding standard deviations (the numerical values are presented in Table XX in Appendix).

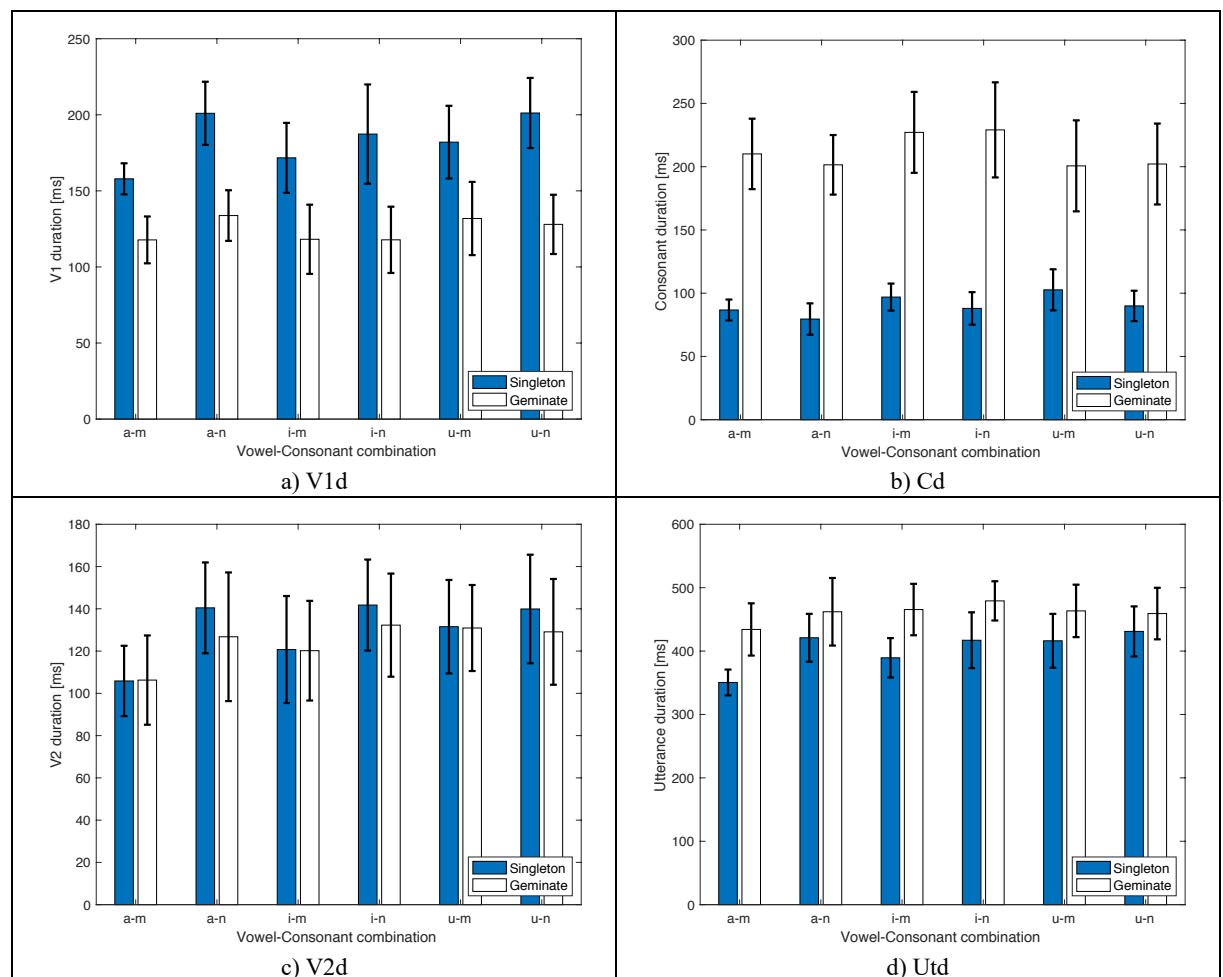

**Figure 4** – Average and standard deviation of time domain parameters for words containing nasals in singleton vs. geminate forms, averaged over all repetitions and speakers (all values are expressed in milliseconds).

Generally speaking, results in Figure 4 highlight a general tendency to shorten the pre-consonant vowel duration V1d and lengthen consonant duration Cd in geminate vs. singleton words, while the post-consonant vowel duration V2d does not appear to be affected by gemination in a systematic way. Geminate words were in average – over all words – about 14% longer than singletons. A detailed statistical analysis follows.

A repeated measurements ANOVA test was performed on female and male speakers data separately, averaged over repetitions. Form (singleton vs. geminate) was used as a between-subjects factor, while Vowel ([a, i, u]) and Consonant ([m, n]) were considered as within-subject factors. Note that the distinction between Form as a between-subjects factor vs. Vowel and Consonant as within-subject factors is not related to the way data were collected, since each speaker recorded all combinations of Form, Vowel and Consonant. The distinction was rather the result of an experiment design choice to highlight the impact of gemination. For each parameter, Table II contains the test variable F and the corresponding p value for each factor and for the interaction between each within-subjects factor and the between-subjects factor; bold values indicate significant values, with threshold set as $p^*=0.05$.

|  |  | Female | | Male | |
| --- | --- | --- | --- | --- | --- |
|  |  | **F** | **p** | **F** | **p** |
| **V1d** | **Form** | **18.03** | **0.013** | **23.98** | **0.008** |
|  | Vowel*Form | 0.012 | 0.919 | .554 | .595 |
|  | Consonant*Form | 2.597 | 0.182 | **9.857** | **0.035** |
|  | Vowel | 2.254 | 0.208 | 1.817 | .224 |
|  | Consonant | **9.011** | **0.040** | **11.293** | **0.028** |
| **Cd** | **Form** | **45.915** | **0.002** | **78.946** | **0.001** |
|  | **Vowel*Form** | **21.006** | **0.001** | **5.658** | **0.029** |
|  | Consonant*Form | **11.863** | **0.026** | 0.187 | 0.688 |
|  | Vowel | **8.042** | **0.012** | **8.038** | **0.012** |
|  | Consonant | 0.544 | 0.502 | 4.119 | 0.112 |
| **V2d** | Form | 0.039 | 0.852 | 0.181 | 0.692 |
|  | Vowel*Form | 1.271 | 0.332 | 1.029 | 0.400 |
|  | Consonant*Form | 2.75 | 0.173 | 1.218 | 0.332 |
|  | Vowel | **4.626** | **0.046** | **12.203** | **0.004** |
|  | Consonant | **19.568** | **0.011** | **11.668** | **0.027** |
| **Utd** | **Form** | **14.776** | **0.018** | 2.257 | 0.207 |
|  | Vowel*Form | 1.845 | 0.219 | **4.668** | **0.045** |
|  | Consonant*Form | 0.235 | 0.653 | 5.983 | 0.071 |
|  | Vowel | **4.845** | **0.042** | **10.121** | **0.006** |
|  | Consonant | **10.405** | **0.032** | 5.845 | 0.073 |

**Table II** – Results of the repeated measurements multivariate ANOVA test performed on time domain parameters for words containing nasals on female and male speakers separately, averaging data over repetitions; test variable F and corresponding probability p at which the null hypothesis can be rejected are presented for the between-subjects factor Form (singleton vs. geminate), for the within-subjects factors Vowel ([a, i, u]) and Consonant ([m, n]), and for their interactions; bold characters indicate significantly different values, with threshold set as $p^*=0.05$.

Table II shows that gemination has a significant impact on the average value of Cd and V1d for both female and male speakers, and of Utd for female speakers. No significant variations were observed for V2d.

Vowel has a significant impact on the Cd parameter for both female and male speakers; the same behavior can be observed for V2d and Utd. As for the Consonant factor, significant variations can be observed for V1d and V2d for both female and male speakers, and for Utd for female speakers.

In order to get further insight on the impact of gemination, additional univariate ANOVA tests were carried out separately for each vowel and consonant, considering Form as the only fixed factor. Male and female speakers were in this case combined, since Table II highlighted no major differences for the two genders with respect to gemination. Results are presented in Table III, showing the test variable F and corresponding probability p of validity of the null hypothesis; values in bold indicate statistically significant variations between singleton vs. geminate groups, with threshold set as $p^*=0.05$.

Results of Table III confirm that Cd and V1d are both impacted by gemination; variations of both parameters between singletons and geminates groups were in fact significant for all combinations of consonants and vowels. A weaker significance was observed for Utd, with significant variations in all cases but with markedly larger p values. Finally, the post-consonant vowel duration V2d did not vary significantly between singletons vs. geminates for any combination of vowels and consonants.

|   |       | a      |        |        |        | i      |        |        |        | u      |        |        |        |
|---|-------|--------|--------|--------|--------|--------|--------|--------|--------|--------|--------|--------|--------|
|   |       | V1d    | Cd     | V2d    | Utd    | V1d    | Cd     | V2d    | Utd    | V1d    | Cd     | V2d    | Utd    |
| m | F(1,34) | 84.98  | 324.43 | 0      | 59.67  | 49.34  | 268.26 | 0      | 39.92  | 39.4   | 111.01 | 0.01   | 11.4   |
|   | p     | 8.98E-11 | 5.8E-19 | 0.9487 | 5.52E-09 | 4.15E-08 | 1.06E-17 | 0.9472 | 3.33E-07 | 3.77E-07 | 3.01E-12 | 0.9299 | 0.0019 |
| n | F(1,34) | 114.91 | 377.91 | 2.43   | 7.1    | 56.54  | 227.17 | 1.54   | 23.95  | 105.92 | 194.06 | 1.64   | 4.44   |
|   | p     | 1.91E-12 | 5.42E-20 | 0.1284 | 0.0117 | 9.92E-09 | 1.29E-16 | 0.2237 | 2.35E-05 | 5.55E-12 | 1.3E-15 | 0.2092 | 0.0425 |

**Table III** - Test variable F and corresponding probability p at which the null hypothesis can be rejected obtained in the univariate ANOVA test performed on time domain parameters for words containing nasals using the Form (singleton vs. geminate) as fixed factor, for each combination of consonants [m, n] and vowels [a, i, u]; bold characters indicate significantly different values, with threshold set as p*=0.05.

Next, a Spearman Rank correlation coefficient test was carried out in order to verify whether any correlation between time domain parameters could be identified in relation to gemination; results are presented in Table IVa) for singleton and geminated words separately, and in Table IVb) for all combined words.

|           |        | Singleton |        |        | Geminate |        |        |
|-----------|--------|-----------|--------|--------|----------|--------|--------|
|           |        | V1d s.    | Cd s.  | V2d s. | V1d g.   | Cd g.  | V2d g. |
| Singleton | V1d s. | 1.00      | -0.15  | 0.45   |          |        |        |
|           | Cd s.  | -0.15     | 1.00   | -0.09  | not significant | | |
|           | V2d s. | 0.45      | -0.09  | 1.00   |          |        |        |
| Geminate  | V1d g. |           |        |        | 1.00     | -0.28  | 0.39   |
|           | Cd g.  | not significant | | | -0.28    | 1.00   | -0.15  |
|           | V2d g. |           |        |        | 0.39     | -0.15  | 1.00   |

|     | V1d   | Cd    | V2d   |
|-----|-------|-------|-------|
| V1d | 1.00  | -0.77 | 0.35  |
| Cd  | -0.77 | 1.00  | -0.17 |
| V2d | 0.35  | -0.17 | 1.00  |

a) Separate groups (singleton vs. geminate)     b) Combined

**Table IV** - Spearman Rank Correlation Coefficient $r_s$ of time domain parameters for words containing singleton and geminate nasals (Table IVa)), and for all words, singleton and geminate combined (Table IVb)). Bold characters indicate significant correlations, with threshold set at p*=0.05.

Note that correlation coefficients close to 0 indicate negligible correlation between parameters, positive coefficients indicate direct correlation, and negative coefficients indicate inverse correlation. Table IV shows that a strong inverse correlation is present for V1d vs. Cd in the combined group, while a weaker one can be observed for the group of geminated words; no correlation was observed for V1d vs. Cd for singleton words. All groups are characterized by a significant positive correlation between V1d and V2d suggesting that a weak compensation for the lengthening of the consonant involves V2d as well. However, Cd vs. V2d negative correlation was weaker, and only significant in the combined group. A test based on Pearson's correlation highlighted a good match between coefficients obtained with the two tests, suggesting the existence of linear relationship between parameters.

### 4.1.2. Results in the frequency domain

Tables V and VI show the mean and standard deviation of frequency domain parameters, for female vs. male speakers, singleton vs. geminate forms, and for each vowel, in reference frames: 1) V1 CENTER, 2) V1 OFFSET, 3) V1-TO-C TRANSITION (Table V) and 4) C ONSET, 5) C CENTER, 6) C OFFSET, 7) V2 ONSET, 8) V2 CENTER (Table VI). Values in both tables are averaged over consonants, speakers and repetitions.

Results in Tables V and VI show an increased F0 average in geminate words for male speakers, in all frames, while no clear effect of gemination can be observed on pitch for female speakers, and on formants for neither group of speakers. A detailed statistical analysis is provided in the following.

|   |   |   | V1 CENTER ||||||||
|   |   |   | Female (Hz) |||| Male (Hz) ||||
|   |   |   | F0 | F1 | F2 | F3 | F0 | F1 | F2 | F3 |
|---|---|---|---|---|---|---|---|---|---|---|
| a | Singleton | Mean | 189 | 1000 | 1488 | 3064 | 114 | 835 | 1329 | 2571 |
|   |   | StD | 43 | 100 | 71 | 157 | 6 | 29 | 70 | 201 |
|   | Geminate | Mean | 188 | 1028 | 1582 | 3020 | 119 | 831 | 1348 | 2769 |
|   |   | StD | 41 | 92 | 109 | 187 | 8 | 25 | 53 | 238 |
| i | Singleton | Mean | 197 | 390 | 2786 | 3565 | 127 | 277 | 2295 | 3220 |
|   |   | StD | 44 | 97 | 174 | 390 | 8 | 27 | 28 | 142 |
|   | Geminate | Mean | 201 | 399 | 2776 | 3559 | 134 | 280 | 2306 | 3237 |
|   |   | StD | 41 | 84 | 150 | 423 | 7 | 7 | 25 | 112 |
| u | Singleton | Mean | 198 | 405 | 705 | 2913 | 133 | 310 | 594 | 2415 |
|   |   | StD | 85 | 171 | 48 | 249 | 6 | 31 | 39 | 91 |
|   | Geminate | Mean | 206 | 418 | 742 | 3135 | 143 | 295 | 625 | 2377 |
|   |   | StD | 40 | 71 | 33 | 283 | 8 | 21 | 51 | 51 |
|   |   |   | V1 OFFSET ||||||||
|   |   |   | Female (Hz) |||| Male (Hz) ||||
|   |   |   | F0 | F1 | F2 | F3 | F0 | F1 | F2 | F3 |
| a | Singleton | Mean | 178 | 913 | 1499 | 3098 | 108 | 790 | 1332 | 2536 |
|   |   | StD | 43 | 55 | 153 | 210 | 6 | 38 | 131 | 208 |
|   | Geminate | Mean | 185 | 983 | 1549 | 3048 | 118 | 809 | 1312 | 2693 |
|   |   | StD | 42 | 65 | 161 | 216 | 10 | 28 | 142 | 289 |
| i | Singleton | Mean | 184 | 377 | 2776 | 3528 | 119 | 297 | 2317 | 3251 |
|   |   | StD | 46 | 87 | 162 | 406 | 8 | 19 | 47 | 137 |
|   | Geminate | Mean | 195 | 390 | 2769 | 3499 | 133 | 280 | 2306 | 3144 |
|   |   | StD | 44 | 86 | 178 | 483 | 6 | 16 | 15 | 201 |
| u | Singleton | Mean | 182 | 368 | 801 | 2973 | 122 | 316 | 679 | 2382 |
|   |   | StD | 79 | 158 | 99 | 182 | 7 | 27 | 133 | 102 |
|   | Geminate | Mean | 199 | 405 | 783 | 2971 | 138 | 288 | 714 | 2347 |
|   |   | StD | 40 | 80 | 158 | 80 | 9 | 13 | 149 | 71 |
|   |   |   | V1-TO-C TRANSITION ||||||||
|   |   |   | Female (Hz) |||| Male (Hz) ||||
|   |   |   | F0 | F1 | F2 | F3 | F0 | F1 | F2 | F3 |
| a | Singleton | Mean | 178 | 883 | 1514 | 3083 | 107 | 850 | 1283 | 2532 |
|   |   | StD | 42 | 34 | 248 | 202 | 6 | 26 | 176 | 193 |
|   | Geminate | Mean | 184 | 957 | 1519 | 3057 | 116 | 833 | 1278 | 2630 |
|   |   | StD | 41 | 61 | 217 | 220 | 10 | 47 | 150 | 310 |
| i | Singleton | Mean | 181 | 355 | 2756 | 3524 | 116 | 295 | 2332 | 3251 |
|   |   | StD | 43 | 88 | 170 | 416 | 9 | 20 | 47 | 176 |
|   | Geminate | Mean | 193 | 381 | 2745 | 3565 | 132 | 282 | 2317 | 3146 |
|   |   | StD | 43 | 88 | 217 | 505 | 7 | 16 | 48 | 174 |
| u | Singleton | Mean | 178 | 345 | 790 | 2986 | 120 | 310 | 712 | 2427 |
|   |   | StD | 77 | 142 | 133 | 149 | 8 | 24 | 184 | 144 |
|   | Geminate | Mean | 195 | 390 | 781 | 3011 | 135 | 282 | 714 | 2330 |
|   |   | StD | 40 | 72 | 113 | 154 | 10 | 18 | 154 | 120 |

**Table V** – Average and standard deviation of pitch F0 and formants F1, F2 and F3 in reference frames V1 CENTER, V1 OFFSET and V1-TO-C TRANSITION for words containing nasals, for female vs. male speakers, averaged over repetitions, speakers and consonants (frequencies are in Hz).

|  |  |  | C ONSET / C CENTER / C OFFSET ||||||
|  |  |  | Female (Hz) ||| Male (Hz) |||
|  |  |  | F0 | F0 | F0 | F0 | F0 | F0 |
| a | Singleton | Mean | 172 | 168 | 162 | 106 | 105 | 104 |
|  |  | StD | 39 | 34 | 29 | 6 | 5 | 6 |
|  | Geminate | Mean | 177 | 164 | 158 | 115 | 107 | 106 |
|  |  | StD | 39 | 32 | 26 | 11 | 13 | 15 |
| i | Singleton | Mean | 176 | 169 | 167 | 114 | 111 | 110 |
|  |  | StD | 41 | 35 | 33 | 10 | 11 | 11 |
|  | Geminate | Mean | 190 | 173 | 163 | 129 | 115 | 114 |
|  |  | StD | 42 | 37 | 29 | 6 | 11 | 12 |
| u | Singleton | Mean | 173 | 170 | 163 | 118 | 116 | 114 |
|  |  | StD | 74 | 71 | 25 | 9 | 9 | 9 |
|  | Geminate | Mean | 191 | 176 | 168 | 131 | 117 | 115 |
|  |  | StD | 39 | 33 | 27 | 11 | 12 | 13 |
|  |  |  | V2 ONSET ||||||||
|  |  |  | Female (Hz) |||| Male (Hz) ||||
|  |  |  | F0 | F1 | F2 | F3 | F0 | F1 | F2 | F3 |
| a | Singleton | Mean | 158 | 887 | 1516 | 3118 | 104 | 801 | 1343 | 2521 |
|  |  | StD | 26 | 30 | 224 | 169 | 7 | 46 | 192 | 176 |
|  | Geminate | Mean | 155 | 911 | 1523 | 3094 | 106 | 818 | 1349 | 2788 |
|  |  | StD | 23 | 53 | 168 | 178 | 17 | 52 | 155 | 392 |
| i | Singleton | Mean | 163 | 321 | 2843 | 3520 | 110 | 271 | 2371 | 3244 |
|  |  | StD | 31 | 71 | 152 | 344 | 10 | 25 | 69 | 164 |
|  | Geminate | Mean | 163 | 319 | 2773 | 3531 | 113 | 301 | 2397 | 3192 |
|  |  | StD | 27 | 57 | 217 | 399 | 12 | 17 | 68 | 202 |
| u | Singleton | Mean | 160 | 321 | 848 | 3005 | 113 | 314 | 739 | 2423 |
|  |  | StD | 64 | 131 | 120 | 147 | 9 | 15 | 214 | 169 |
|  | Geminate | Mean | 167 | 336 | 842 | 3222 | 115 | 316 | 771 | 2428 |
|  |  | StD | 26 | 60 | 183 | 448 | 14 | 18 | 231 | 122 |
|  |  |  | V2 CENTER ||||||||
|  |  |  | Female (Hz) |||| Male (Hz) ||||
|  |  |  | F0 | F1 | F2 | F3 | F0 | F1 | F2 | F3 |
| a | Singleton | Mean | 153 | 909 | 1525 | 3198 | 107 | 816 | 1347 | 2519 |
|  |  | StD | 21 | 52 | 120 | 212 | 9 | 27 | 105 | 169 |
|  | Geminate | Mean | 151 | 954 | 1538 | 3109 | 107 | 805 | 1358 | 2808 |
|  |  | StD | 17 | 90 | 104 | 166 | 16 | 55 | 116 | 490 |
| i | Singleton | Mean | 158 | 314 | 2821 | 3535 | 109 | 280 | 2382 | 3198 |
|  |  | StD | 27 | 59 | 159 | 355 | 12 | 14 | 116 | 177 |
|  | Geminate | Mean | 160 | 319 | 2801 | 3524 | 112 | 299 | 2391 | 3163 |
|  |  | StD | 26 | 57 | 184 | 409 | 14 | 18 | 102 | 228 |
| u | Singleton | Mean | 153 | 308 | 824 | 3007 | 113 | 312 | 670 | 2469 |
|  |  | StD | 60 | 124 | 110 | 181 | 9 | 25 | 175 | 160 |
|  | Geminate | Mean | 162 | 329 | 781 | 3258 | 115 | 308 | 683 | 2369 |
|  |  | StD | 25 | 67 | 45 | 557 | 18 | 20 | 137 | 27 |

**Table VI** - Average and standard deviation of pitch F0 and formants F1, F2 and F3 in reference frames V2 ONSET and V2 CENTER, and of pitch F0 in reference frames C1 ONSET, C CENTER and C OFFSET for words containing nasals, for female vs. male speakers, averaged with respect to repetitions, speakers and consonants (frequencies are in Hz).

A multi-factor univariate ANOVA test was carried using Form, Vowel and Consonant as fixed factors on female vs. male speakers. Results are presented in Table VII, that shows a factor vs. parameter matrix: a checked cell at the intersection between a factor and a parameter indicates a significant difference in the average value of the parameter due to that factor. Results in Table VII indicate that Form does not cause significant differences for any of the frequency domain parameter for female speakers, while, for male speakers, F0 shows a significant difference in the three frames related to the first vowel as well as in the C ONSET frame. Vowel induced, as expected, significant differences in both F0 (intrinsic pitch) for V1-related frames, and in formants F1, F2 and F3, in frames related to both V1 and V2. Factor Consonant led to significant differences only in sporadic cases, in particular in frames V1 OFFSET, V1-TO-C TRANSITION, V2 ONSET and V2 CENTER, and only for F2. Overall, in nasals frequency domain parameters do not seem to provide cues to gemination across speakers of different genders.

|  |  | Female | | | | Male | | | |
| --- | --- | --- | --- | --- | --- | --- | --- | --- | --- |
|  |  | F0 | F1 | F2 | F3 | F0 | F1 | F2 | F3 |
| V1 CENTER | **Form** |  |  |  |  | X |  |  |  |
|  | **Vowel** |  | X | X | X | X | X | X | X |
|  | **Consonant** |  |  |  |  |  |  |  |  |
| V1 OFFSET | **Form** |  |  |  |  | X |  |  |  |
|  | **Vowel** |  | X | X | X | X | X | X | X |
|  | **Consonant** |  |  | X |  |  |  | X |  |
| V1-TO-C TRANSITION | **Form** |  |  |  |  | X | X |  |  |
|  | **Vowel** |  | X | X | X | X | X | X | X |
|  | **Consonant** |  |  |  |  |  |  | X |  |
| C ONSET | **Form** |  |  |  |  | X |  |  |  |
|  | **Vowel** |  | N/A | | | X | N/A | | |
|  | **Consonant** |  |  |  |  | X |  |  |  |
| C CENTER | **Form** |  |  |  |  |  |  |  |  |
|  | **Vowel** |  | N/A | | | | N/A | | |
|  | **Consonant** |  |  |  |  |  |  |  |  |
| C OFFSET | **Form** |  |  |  |  |  |  |  |  |
|  | **Vowel** |  | N/A | | | | N/A | | |
|  | **Consonant** |  |  |  |  |  |  |  |  |
| V2 ONSET | **Form** |  |  |  |  |  |  |  |  |
|  | **Vowel** |  | X | X | X |  | X | X | X |
|  | **Consonant** |  |  |  |  |  |  | X |  |
| V2 CENTER | **Form** |  |  |  |  |  |  |  |  |
|  | **Vowel** |  | X | X | X |  | X | X | X |
|  | **Consonant** |  |  |  |  |  |  | X |  |

**Table VII** – Results of the multi-factor univariate ANOVA test performed on frequency domain parameters in vowel reference frames V1 CENTER, V1 OFFSET, V1-TO-C TRANSITION, C ONSET, C CENTER, C OFFSET, V2 ONSET and V2 CENTER for words containing nasals using Form, Vowel and Consonant as fixed factors; a checked cell at the intersection between a parameter and a factor indicates a significant difference between average values for the parameter with respect to the factor.

### 4.1.3. Results in the energy domain

Figure 5 shows the average values of energy domain parameters (for a list of parameters refer to Section 3.1.5; the numerical values are presented in Table XXI in Appendix). Since in the case of energy domain parameters the impact of gender was not expected to be as strong as for frequency domain parameters, results are presented here averaged over all speakers and repetitions.

Average values presented in Figure 5 do not indicate any clear trend. A statistical multi-factor univariate ANOVA test was thus performed in order to determine if statistically significative differences between averages exist. The test considered the fixed factors Form, Vowel, Consonant and Gender, and was applied to all words combined. Results of the ANOVA test are presented in Table VIII as a matrix of factor vs. parameter in which a checked cell indicates a significant difference in the average value of the parameter due to factor. Table IX shows that $E_{totC}$ varies significantly with Form (gemination). As for the other factors, Consonant led to significant differences for all parameters while Vowel led to significant variations for all parameters related to V1 except for $E_{iV1-C}$ (energy of transition frame from vowel to consonant). Finally, the Gender factor led to no significant variations.

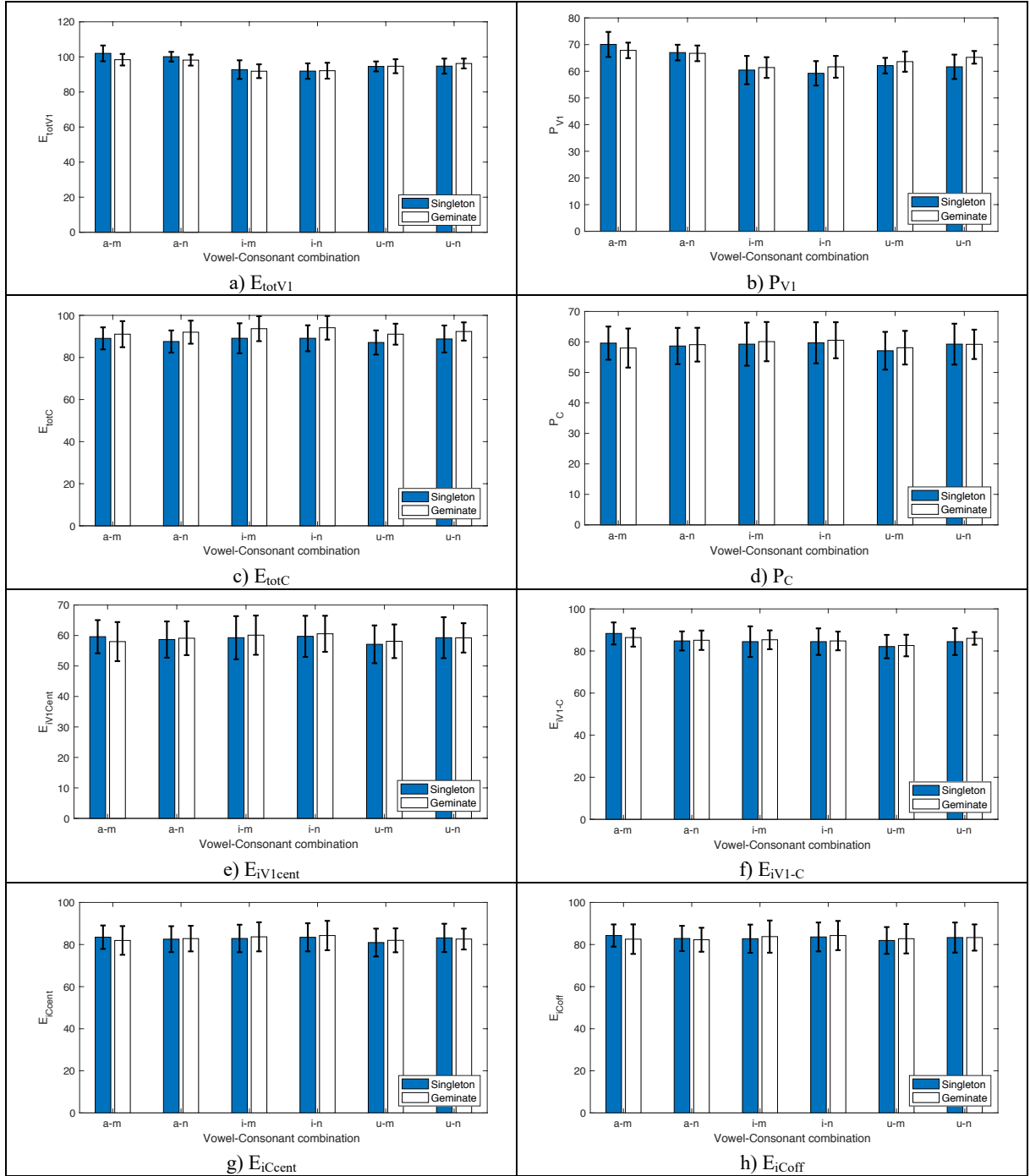

**Figure 5** – Average and standard deviation of energy domain parameters for each combination of consonants [m, n], vowels [a, i, u] and singleton vs. geminate form, averaged over repetitions and speakers (values are in logarithmic form; for a list of parameters refer to Section 3.1.5).

|  | $E_{totV1}$ | $P_{mV1}$ | $E_{totC}$ | $P_{mC}$ | $E_{iV1cent}$ | $E_{iV1-C}$ | $E_{iCcent}$ | $E_{iCoffset}$ |
|---|---|---|---|---|---|---|---|---|
| **Form** |  |  | X |  |  |  |  |  |
| **Vowel** | X | X |  |  | X |  |  |  |
| **Consonant** | X | X | X | X | X | X | X | X |
| **Gender** |  |  |  |  |  |  |  |  |

**Table VIII** - Results of the multi-factor univariate ANOVA test performed on energy domain parameters using Form, Vowel, Consonant and Gender for all words containing nasals; a checked cell indicates a significant difference between average values for the parameter with respect to the factor.

## 4.2. Results on liquids

### 4.2.1. Results in the time domain

The time domain parameters listed in Section 3.1.3 were computed for each of the 108 singleton and 108 geminate words containing liquids.

Results are presented in Figure 6, showing the average values and standard deviations of V1d, Cd, V2d and Utd for all combinations of vowels [a, i, u] and consonants [r, l] in geminate vs. singleton forms, averaged over all repetitions and speakers deviations (numerical values are presented in Table XXII in Appendix).

Figure 6 shows that, as regards V1d and Cd, liquids behave somewhat like nasals; V1d tends to decrease with gemination, while the opposite is true for Cd. A trend to become shorter in presence of gemination can be also observed for V2d (with the exception of ili vs. illi) and Utd. A statistical analysis was carried out to investigate whether the trends are statistically significant.

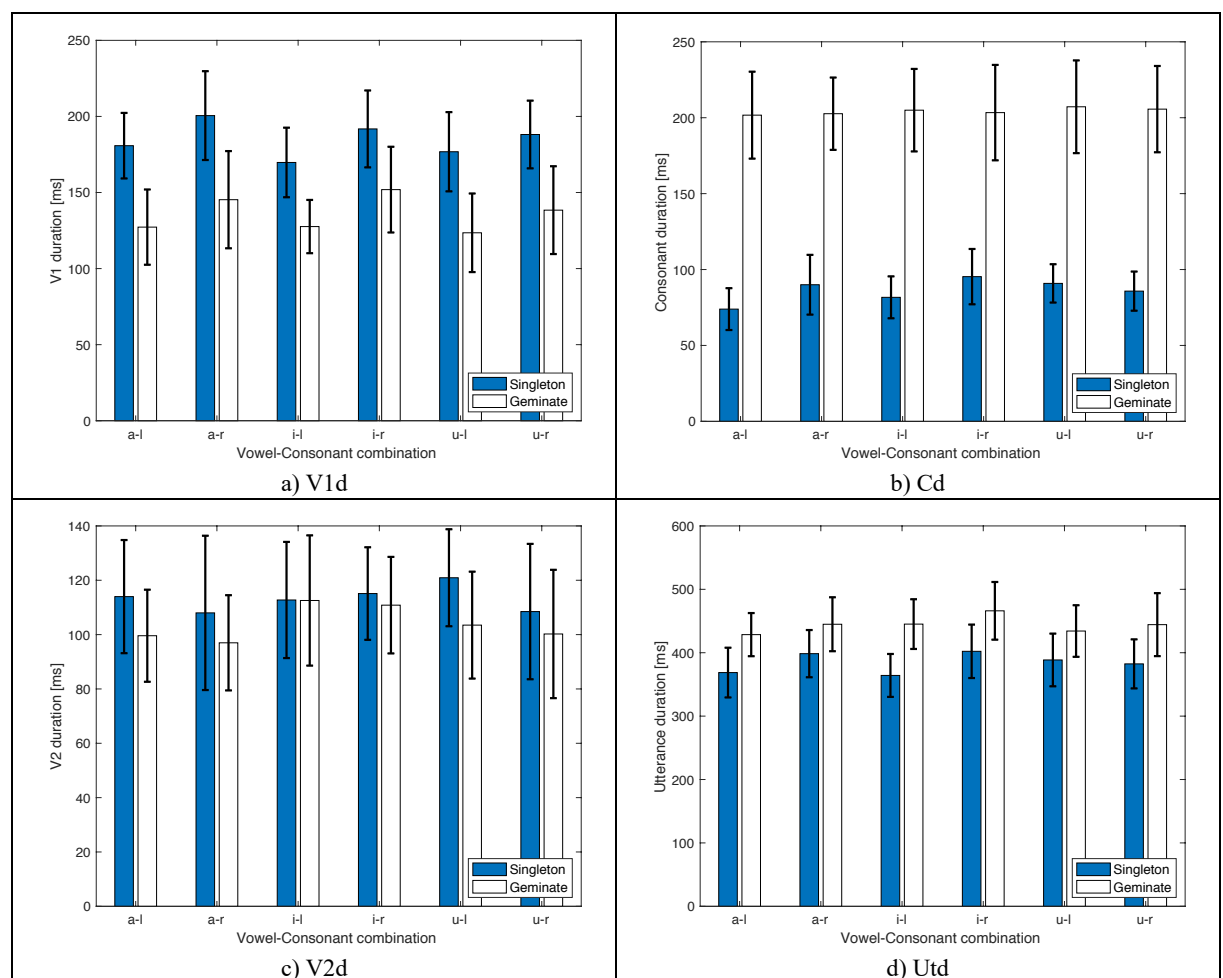

**Figure 6** – Average and standard deviation of time domain parameters for words containing liquids in singleton vs. geminate forms, averaged over all repetitions and speakers (all values are expressed in milliseconds).

A repeated measurements ANOVA test was performed on female and male speakers data separately, after averaging over repetitions, using Form (singleton vs. geminate) as a between-subjects factor, and Vowel [a, i, u] and Consonant [l, r] as within-subjects factors. Results are presented in Table IX, that shows, for each parameter, the test variable F and corresponding p value for each factor and for the interaction between each within-subjects factor and the between-subjects factor. Bold values indicate significant variations, with threshold set as $p^*=0.05$.

|  |  | Female |  | Male |  |
| --- | --- | --- | --- | --- | --- |
|  |  | F | p | F | p |
| V1d | Form | 3.483 | 0.135 | **154.8** | **<0.001** |
|  | Vowel*Form | 2.393 | 0.153 | 0.252 | 0.783 |
|  | Consonant*Form | 0.348 | 0.587 | 0.213 | 0.669 |
|  | Vowel | 0.820 | 0.474 | 1.292 | 0.326 |
|  | Consonant | 5.200 | 0.085 | **42.250** | **0.003** |
| Cd | Form | **71.124** | **0.001** | 500.170 | <0.001 |
|  | Vowel*Form | 0.201 | 0.822 | 0.179 | 0.839 |
|  | Consonant*Form | 0.022 | 0.890 | **8.287** | **0.045** |
|  | Vowel | 0.211 | 0.814 | 1.155 | 0.353 |
|  | Consonant | 0.300 | 0.613 | 4.409 | 0.104 |
| V2d | Form | 1.498 | 0.288 | 0.175 | 0.698 |
|  | Vowel*Form | 0.977 | 0.417 | 2.256 | 0.167 |
|  | Consonant*Form | 1.046 | 0.364 | 0.098 | 0.770 |
|  | Vowel | 0.469 | 0.642 | **8.569** | **0.01** |
|  | Consonant | 0.049 | 0-836 | 6.005 | 0.07 |
| Utd | Form | 4.711 | 0.096 | **13.417** | **0.022** |
|  | Vowel*Form | 1.621 | 0.256 | 1.022 | 0.402 |
|  | Consonant*Form | 0.937 | 0.388 | 3.532 | 0.133 |
|  | Vowel | 0.001 | 0.999 | **5.079** | **0.038** |
|  | Consonant | 3.373 | 0.140 | **15.997** | **0.016** |

**Table IX** – Results of the repeated measurements ANOVA test for liquids performed on time domain parameters, for female vs. male speakers separately. Values are averaged over all repetitions. Test variable F and corresponding probability p at which the null hypothesis can be rejected are presented for the between-subjects factor Form (singleton vs. geminate), for the within-subjects factors Vowel [a, i, u] and Consonant [l, r], and for their interactions. Bold characters indicate significant variations, with threshold set as p*=0.05.

In terms of gemination, results in Table IX highlight a significant variation of Cd for both female and male speakers, while only male speakers show a significant variation of both V1d and Utd. No significant variations were observed for V2d.

As for other factors, Consonant has a significant impact on V1d and Utd for male speakers. Finally, Vowel was significant only for V2d and male speakers.

As for nasals (see Section 4.1.1), additional univariate ANOVA tests for the Form factor (gemination) were carried out for each combination of vowel and consonant separately, on combined female and male speakers data. Results are presented in Table X, and confirm the combined results presented in Table IX. Consonant duration Cd is strongly affected by gemination for all combinations of vowels and consonants. Gemination also has an impact on V1d in all cases, albeit with larger p values, and on Utd with an even weaker significance. As a side note, a significant variation for V2d was observed but only for [l] uttered within [a] and [u].

|  |  | a | | | | i | | | | u | | | |
| --- | --- | --- | --- | --- | --- | --- | --- | --- | --- | --- | --- | --- | --- |
|  |  | V1d | Cd | V2d | Utd | V1d | Cd | V2d | Utd | V1d | Cd | V2d | Utd |
| l | F(1,34) | 47.95 | 290.93 | 5.18 | 23.89 | 38.53 | 294.48 | 0 | 43.97 | 37.98 | 223.05 | 7.75 | 11.08 |
|  | p | 5.58E-08 | 3.09E-18 | 0.0293 | 2.4€-05 | 4.64E-07 | 2.57E-18 | 0.98 | 1.32E-07 | 5.3E-07 | 1.69E-16 | 0.0087 | 0.0021 |
| r | F(1,34) | 29.37 | 239.31 | 1.96 | 12.11 | 20 | 159.21 | 0.54 | 19.09 | 33.52 | 265.51 | 1.04 | 17.4 |
|  | p | 4.9E-06 | 5.91E-17 | 0.1705 | 0.0014 | 8.22E-05 | 2.21E-14 | 0.4665 | 0.0001 | 1.61E-06 | 1.24E-17 | 0.3153 | 0.0002 |

**Table X** – Test variable F and corresponding probability p at which the null hypothesis can be rejected obtained in the univariate ANOVA test performed on time domain parameters for words containing liquids using Form (singleton vs. geminate) as fixed factor, for each combination of consonants [l, r] and vowels [a, i, u]. Bold characters indicate significantly different values, with threshold set as p*=0.05.

Finally, the Spearman Rank Correlation Coefficient $r_s$ was evaluated, for both singleton and geminate group, first separately and then combined. Results are presented in Table XIa) and Table XIb).

|  |  | **Singleton** | | | **Geminate** | | |
|---|---|---|---|---|---|---|---|
|  |  | V1d s. | Cd s. | V2d s. | V1d g. | Cd g. | V2d g. |
| **Singleton** | V1d s. | 1.00 | 0.1 | 0.24 | | | |
|  | Cd s. | 0.1 | 1.00 | -0.26 | | not significant | |
|  | V2d s. | 0.24 | -0.26 | 1.00 | | | |
| **Geminate** | V1d g. | | | | 1.00 | -0.35 | 0.47 |
|  | Cd g. | | not significant | | -0.35 | 1.00 | -0.28 |
|  | V2d g. | | | | 0.47 | -0.28 | 1.00 |

|  | V1d | Cd | V2d |
|---|---|---|---|
| V1d | 1.00 | -0.64 | 0.38 |
| Cd | -0.64 | 1.00 | -0.32 |
| V2d | 0.38 | -0.32 | 1.00 |

a) Separate groups (singleton vs. geminate)     b) Combined

**Table XI** – Spearman Rank Correlation Coefficient $r_s$ of time domain parameters for singleton and geminate liquid words separately (Table XIa)), and on all words combined (Table XIb)). Bold characters indicate significant correlations, with threshold set at p*=0.05.

Table XIa) shows that results for liquids are in good agreement with those obtained for nasals (Table IV): V1d and Cd are not correlated in singleton words, while a moderate inverse correlation appears in geminate words, and a strong one is observed for the group including all combined words. A Pearson's correlation test led to similar results, indicating that the relationships between parameters are monotonic and linear.

### 4.2.2. Results in the frequency domain

The analysis in the frequency domain of liquids regarded frequency domain parameters F0, F1, F2 and F3 for both vowel related reference frames, and pitch F0 for consonant related frames, as defined in Section 3.1.2.

The average value and standard deviations of F0, F1, F2 and F3 in V1 CENTER, V1 OFFSET and V1-TO-C TRANSITION reference frames are presented in Table XII, while Table XIII presents average value and standard deviation of F0 in C ONSET, C CENTER and C OFFSET reference frames and of F0, F1, F2 and F3 in V2 ONSET and V2 CENTER reference frames. Data were obtained for female vs. male speakers separately and for each combination of vowels [a, i, u] and forms (singleton vs. geminate), averaged over all speakers, consonants and repetitions.

|  |  |  | **V1 CENTER** | | | | | | | |
|---|---|---|---|---|---|---|---|---|---|---|
|  |  |  | **Female (Hz)** | | | | **Male (Hz)** | | | |
|  |  |  | F0 | F1 | F2 | F3 | F0 | F1 | F2 | F3 |
| **a** | Singleton | Mean | 195 | 1050 | 1540 | 2902 | 116 | 769 | 1298 | 2489 |
|  |  | StD | 28 | 79 | 120 | 182 | 3 | 10 | 58 | 98 |
|  | Geminate | Mean | 194 | 1025 | 1546 | 2894 | 122 | 774 | 1304 | 2538 |
|  |  | StD | 30 | 57 | 122 | 197 | 9 | 25 | 65 | 136 |
| **i** | Singleton | Mean | 216 | 414 | 2770 | 3511 | 184 | 295 | 2237 | 2956 |
|  |  | StD | 26 | 54 | 107 | 181 | 41 | 15 | 60 | 133 |
|  | Geminate | Mean | 220 | 411 | 2716 | 3355 | 196 | 314 | 2145 | 2813 |
|  |  | StD | 29 | 62 | 106 | 176 | 33 | 28 | 55 | 180 |
| **u** | Singleton | Mean | 223 | 431 | 761 | 2866 | 180 | 335 | 714 | 2361 |
|  |  | StD | 87 | 168 | 53 | 201 | 55 | 30 | 49 | 82 |
|  | Geminate | Mean | 226 | 422 | 778 | 2842 | 172 | 333 | 774 | 2349 |
|  |  | StD | 30 | 23 | 48 | 225 | 41 | 23 | 52 | 122 |
|  |  |  | **V1 OFFSET** | | | | | | | |
|  |  |  | **Female (Hz)** | | | | **Male (Hz)** | | | |
|  |  |  | F0 | F1 | F2 | F3 | F0 | F1 | F2 | F3 |
| **a** | Singleton | Mean | 193 | 1022 | 1573 | 2919 | 112 | 766 | 1306 | 2508 |
|  |  | StD | 26 | 90 | 123 | 234 | 3 | 13 | 72 | 96 |
|  | Geminate | Mean | 197 | 1011 | 1549 | 2903 | 122 | 777 | 1310 | 2532 |
|  |  | StD | 32 | 84 | 113 | 156 | 8 | 26 | 75 | 163 |
| **i** | Singleton | Mean | 209 | 407 | 2769 | 3255 | 160 | 299 | 2250 | 2892 |
|  |  | StD | 25 | 50 | 91 | 518 | 59 | 22 | 64 | 117 |
|  |  | Mean | 212 | 423 | 2675 | 3263 | 178 | 319 | 2131 | 2740 |

|   |   |   | F0 | F1 | F2 | F3 | F0 | F1 | F2 | F3 |
|---|---|---|----|----|----|----|----|----|----|----|
|   | Geminate | StD | 34 | 51 | 160 | 255 | 47 | 34 | 115 | 262 |
| u | Singleton | Mean | 220 | 427 | 794 | 2851 | 179 | 329 | 712 | 2354 |
|   |           | StD  | 86  | 165 | 59  | 206  | 73  | 30  | 48  | 75   |
|   | Geminate  | Mean | 232 | 427 | 801 | 2800 | 179 | 334 | 790 | 2332 |
|   |           | StD  | 28  | 22  | 40  | 218  | 60  | 26  | 77  | 146  |
|   |   |   | colspan=8 V1-TO-C TRANSITION |
|   |   |   | Female (Hz) | | | | Male (Hz) | | | |
|   |   |   | F0 | F1 | F2 | F3 | F0 | F1 | F2 | F3 |
| a | Singleton | Mean | 190 | 950 | 1612 | 2925 | 109 | 716 | 1263 | 2495 |
|   |           | StD  | 26  | 88  | 118  | 223  | 3   | 53  | 159  | 95   |
|   | Geminate  | Mean | 187 | 918 | 1592 | 2907 | 119 | 749 | 1334 | 2495 |
|   |           | StD  | 43  | 64  | 132  | 294  | 9   | 20  | 88   | 127  |
| i | Singleton | Mean | 207 | 411 | 2675 | 3313 | 155 | 321 | 2222 | 2771 |
|   |           | StD  | 23  | 48  | 140  | 210  | 67  | 33  | 75   | 157  |
|   | Geminate  | Mean | 217 | 425 | 2536 | 3101 | 157 | 347 | 2029 | 2621 |
|   |           | StD  | 27  | 50  | 204  | 270  | 39  | 69  | 175  | 195  |
| u | Singleton | Mean | 218 | 425 | 889  | 2661 | 153 | 347 | 798  | 2308 |
|   |           | StD  | 85  | 164 | 65   | 286  | 46  | 27  | 56   | 104  |
|   | Geminate  | Mean | 224 | 416 | 892  | 2684 | 168 | 336 | 879  | 2226 |
|   |           | StD  | 26  | 35  | 42   | 275  | 49  | 32  | 95   | 183  |

**Table XII** - Average and standard deviation of F0, F1, F2 and F3 in reference frames V1 CENTER, V1 OFFSET and V1-TO-C TRANSITION for liquids, for female vs. male speakers, averaged over repetitions, speakers and consonants (values are in Hz).

|   |   |   | C ONSET / C CENTER / C OFFSET | | | | | |
|---|---|---|---|---|---|---|---|---|
|   |   |   | Female (Hz) | | | Male (Hz) | | |
|   |   |   | F0 | F0 | F0 | F0 | F0 | F0 |
| a | Singleton | Mean | 187 | 184 | 187 | 108 | 106 | 108 |
|   |           | StD  | 26  | 23  | 26  | 4   | 9   | 10  |
|   | Geminate  | Mean | 195 | 190 | 189 | 116 | 106 | 100 |
|   |           | StD  | 29  | 24  | 24  | 9   | 16  | 11  |
| i | Singleton | Mean | 206 | 221 | 231 | 148 | 122 | 110 |
|   |           | StD  | 26  | 50  | 48  | 55  | 26  | 14  |
|   | Geminate  | Mean | 208 | 209 | 191 | 130 | 115 | 122 |
|   |           | StD  | 33  | 23  | 25  | 33  | 18  | 47  |
| u | Singleton | Mean | 213 | 205 | 210 | 142 | 112 | 108 |
|   |           | StD  | 85  | 81  | 82  | 44  | 8   | 9   |
|   | Geminate  | Mean | 218 | 208 | 208 | 151 | 122 | 120 |
|   |           | StD  | 26  | 23  | 24  | 45  |     | 26  |
|   |   |   | colspan=8 V2 ONSET |
|   |   |   | Female (Hz) | | | | Male (Hz) | | | |
|   |   |   | F0 | F1 | F2 | F3 | F0 | F1 | F2 | F3 |
| a | Singleton | Mean | 186 | 764 | 1688 | 2912 | 105 | 604 | 1350 | 2497 |
|   |           | StD  | 22  | 85  | 140  | 312  | 10  | 50  | 105  | 90   |
|   | Geminate  | Mean | 187 | 707 | 1582 | 2779 | 103 | 577 | 1287 | 2497 |
|   |           | StD  | 24  | 85  | 94   | 397  | 13  | 79  | 55   | 139  |
| i | Singleton | Mean | 209 | 381 | 2524 | 3140 | 152 | 314 | 1966 | 2549 |
|   |           | StD  | 25  | 41  | 97   | 83   | 75  | 30  | 82   | 98   |
|   | Geminate  | Mean | 197 | 410 | 2432 | 3077 | 126 | 329 | 1901 | 2534 |
|   |           | StD  | 25  | 27  | 185  | 191  | 31  | 42  | 76   | 119  |
| u | Singleton | Mean | 198 | 407 | 985  | 2542 | 107 | 353 | 967  | 2039 |
|   |           | StD  | 79  | 157 | 65   | 227  | 9   | 29  | 54   | 165  |
|   | Geminate  | Mean | 210 | 425 | 1037 | 2226 | 130 | 347 | 998  | 2011 |
|   |           | StD  | 18  | 39  | 92   | 211  | 49  | 30  | 126  | 157  |
|   |   |   | colspan=8 V2 CENTER |
|   |   |   | Female (Hz) | | | | Male (Hz) | | | |
|   |   |   | F0 | F1 | F2 | F3 | F0 | F1 | F2 | F3 |
| a | Singleton | Mean | 187 | 978 | 1562 | 2947 | 106 | 734 | 1357 | 2473 |
|   |           | StD  | 10  | 63  | 97   | 295  | 11  | 26  | 94   | 100  |
|   | Geminate  | Mean | 185 | 942 | 1535 | 2945 | 104 | 735 | 1370 | 2490 |
|   |           | StD  | 19  | 67  | 71   | 346  | 13  | 18  | 98   | 201  |
| i | Singleton | Mean | 211 | 373 | 2751 | 3331 | 164 | 302 | 2205 | 2790 |
|   |           | StD  | 12  | 30  | 96   | 162  | 64  | 26  | 100  | 96   |
|   | Geminate  | Mean | 202 | 388 | 2718 | 3282 | 130 | 321 | 2125 | 2694 |
|   |           | StD  | 15  | 24  | 131  | 188  | 25  | 38  | 119  | 94   |
| u | Singleton | Mean | 199 | 407 | 852  | 2883 | 129 | 332 | 823  | 2263 |
|   |           | StD  | 79  | 157 | 20   | 200  | 43  | 27  | 50   | 187  |
|   | Geminate  | Mean | 202 | 412 | 852  | 2849 | 135 | 346 | 864  | 2285 |
|   |           | StD  | 28  | 34  | 40   | 179  | 36  | 33  | 77   | 222  |

**Table XIII** - Average and standard deviation of pitch F0 and formants F1, F2 and F3 in reference frames V2 ONSET and V2 CENTER, and of pitch F0 in reference frames C1 ONSET, C CENTER and C OFFSET for words containing liquids, for female vs. male speakers, averaged over repetitions, speakers and consonants (frequencies are in Hz)..

A multi-factor univariate ANOVA test was then performed using Form, Vowel and Consonant as fixed factors. Results are presented in Table XIV, where a checked cell indicates a significant difference between average values for the parameter with respect to factor. Table XIV shows that gemination only led to statistically significant variations for frequency domain parameters for female speakers, in particular for F1 and F3 at V1 OFFSET, for F2 at C ONSET, and again for F1 at C OFFSET. In the case of male speakers, gemination never led to significant variations of any parameter in any frame.

|                       |           | Female |    |    |    | Male |    |    |    |
|-----------------------|-----------|--------|----|----|----|------|----|----|----|
|                       |           | F0     | F1 | F2 | F3 | F0   | F1 | F2 | F3 |
| V1 CENTER             | Form      |        |    |    |    |      |    |    |    |
|                       | Vowel     |        | X  | X  | X  | X    | X  | X  | X  |
|                       | Consonant |        |    |    |    |      |    |    |    |
| V1 OFFSET             | Form      |        | X  |    | X  |      |    |    |    |
|                       | Vowel     |        |    |    |    | X    | X  | X  | X  |
|                       | Consonant |        |    |    | X  |      |    |    | X  |
| V1-TO-C TRANSITION    | Form      |        |    |    |    |      |    |    |    |
|                       | Vowel     |        | X  | X  | X  | X    | X  | X  | X  |
|                       | Consonant |        |    |    |    |      | X  |    | X  |
| C ONSET               | Form      |        |    |    |    |      |    |    |    |
|                       | Vowel     |        | N/A |   |    |      | N/A |   |    |
|                       | Consonant |        |    |    |    | X    |    |    |    |
| C CENTER              | Form      |        |    |    |    |      |    |    |    |
|                       | Vowel     |        | N/A |   |    |      | N/A |   |    |
|                       | Consonant |        |    |    |    |      |    |    |    |
| C OFFSET              | Form      |        |    |    |    |      |    |    |    |
|                       | Vowel     |        | N/A |   |    | X    | N/A |   |    |
|                       | Consonant |        |    |    |    | X    |    |    |    |
| V2 ONSET              | Form      |        |    |    |    |      |    |    |    |
|                       | Vowel     |        | X  | X  | X  |      | X  | X  | X  |
|                       | Consonant |        |    |    | X  |      |    |    | X  |
| V2 CENTER             | Form      |        |    |    |    |      |    |    |    |
|                       | Vowel     |        | X  | X  | X  | X    | X  | X  | X  |
|                       | Consonant |        |    |    |    | X    |    |    |    |

**Table XIV** – Results of the multi-factor univariate ANOVA test performed on frequency domain parameters in reference frames defined in Section 3.1.2 for words containing liquids using Form, Vowel and Consonant as fixed factors; a checked cell indicates a significant difference of average values for the parameter with respect to the factor..

Vowel was the only factor leading to significant differences in F1, F2 and F3 for both female and male speakers in most frames, with the exception of F3 at C OFFSET for male speakers and at V1 OFFSET for female speakers. Vowel also led to significant variations of F0 for male speakers in all frames except for C ONSET, C CENTER and V2 ONSET. Consonant led to significant differences in F1 in consonant-related frames, in particular at C ONSET (males only), C CENTER and C OFFSET (both female and male speakers), and sporadically in other parameters for male speakers: F0 at C ONSET, C OFFSET and V2 CENTER, F2 at C CENTER, and F3 at V1 OFFSET, V1-TO-C TRANSITION and V2 ONSET.

As a general comment, data for female speakers showed a lower impact of all factors on each parameter. In particular, F0 was not significantly influenced by any factor in any frame.

### 4.2.3. Results in the energy domain

Figure 7 shows mean values and standard deviations for energy domain parameters for each combination of vowels [a, i, u], consonants [l, r] and forms (singleton vs. geminate), averaged over speakers and repetitions (the numerical values are presented in Table XXIII in Appendix). A direct inspection of data in Figure 7 does not highlight any clear trend for any of the parameters, in particular in relation to the gemination.

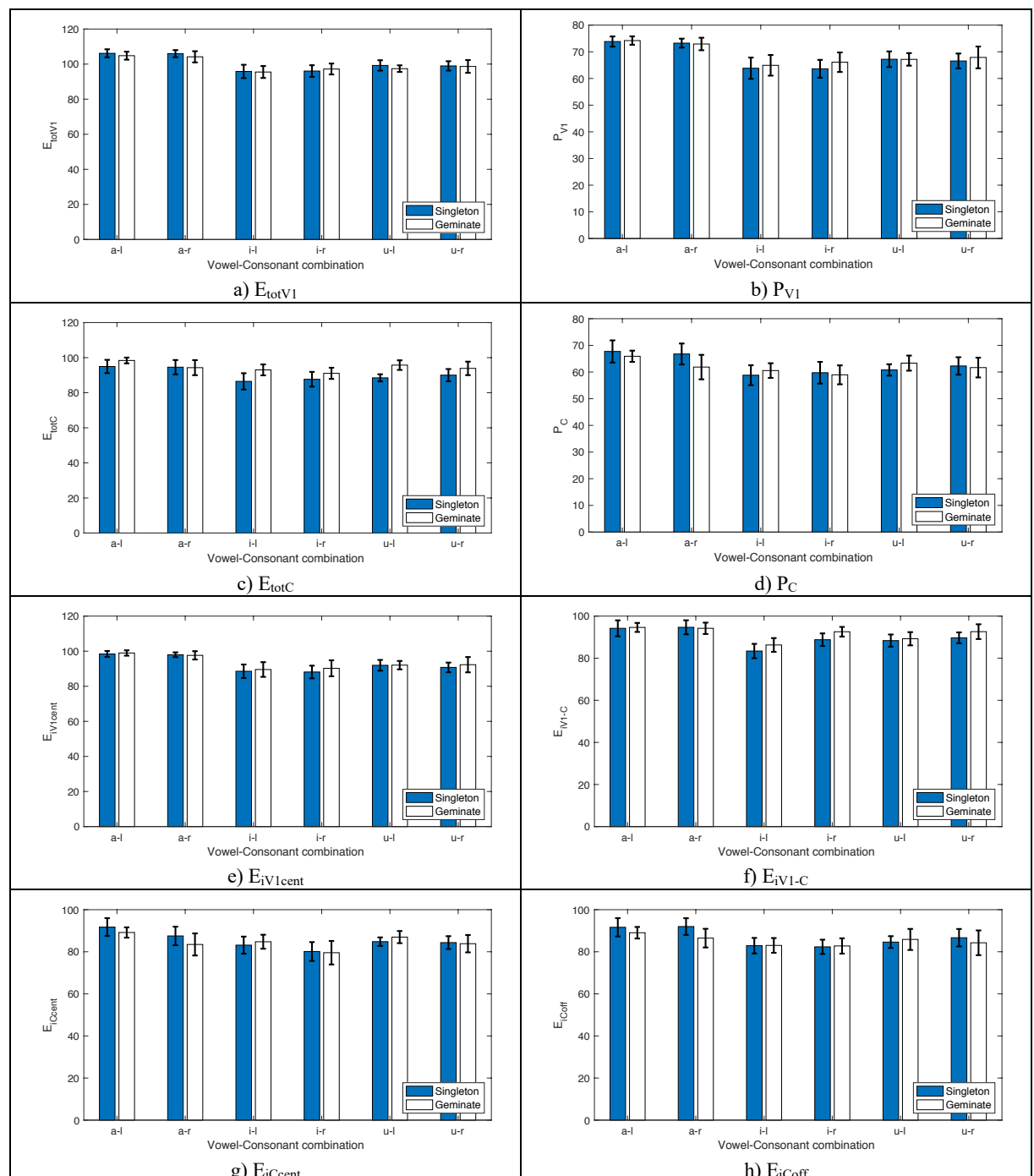

**Figure 7** – Average and standard deviation of energy domain parameters for liquids in singleton vs. geminate forms, averaged over speakers and repetitions (values are in logarithmic form; for a list of parameters refer to Section 3.1.5).

Following the same approach adopted for nasals, a multi-factor univariate ANOVA test considering the fixed factors Form, Vowel, Consonant and Gender was performed over all combined words. Results are presented in Table XV, and show that Form is typically not a significant factor, since only the $E_{totC}$ parameter shows significant variation with gemination.

|  | $E_{totV1}$ | $P_{mV1}$ | $E_{totC}$ | $P_{mC}$ | $E_{iV1cent}$ | $E_{iV1-C}$ | $E_{iCcent}$ | $E_{iCoffset}$ |
|---|---|---|---|---|---|---|---|---|
| **Form** |  |  | X |  |  |  |  |  |
| **Vowel** | X | X | X | X | X | X | X | X |
| **Consonant** |  |  | X |  | X |  |  |  |
| **Gender** |  |  |  |  |  |  | X | X |

**Table XV** - Results of the multi-factor univariate ANOVA test performed for liquids on energy domain parameters using Form, Vowel, Consonant and Gender as fixed factors for all words; a checked cell at the intersection between a parameter and a factor indicates a significant difference between average values for the parameter with respect to the factor.

As for the other factors, Vowel is, by far, the one leading to a stronger impact, since it leads to significant variations of all energy-related parameters. Gender and Consonant only led to sporadic significant differences, respectively for $E_{totC}$ and $E_{iV1cent}$ (Gender) and $E_{iV1-C}$ and $E_{iCcen}$ (Consonant).

## 5. Discussion

### 5.1. Effect of gemination in nasals
Results of the analysis presented in Section 4.1.1 showed a significant increase in consonant duration and a decrease of pre-consonant vowel duration for all combinations of vowels and consonants, and for both female and male speakers. No significant variation was observed in the post-consonant vowel duration. Word duration Utd was only marginally affected by gemination, with significant variations observed for all combinations of vowels with [m], but not with [n], for which only combination with [i] led to significant Utd variations.
In the frequency domain F0 significantly increased when moving from singleton to geminate only for male speakers, and only for reference frames related to V1, in particular in words containing vowels [i] and [u]. No significant variations were observed for formants in any frame for neither female nor male speakers.
Finally, the total energy of the consonant $E_{totC}$ showed significant variations with gemination, while all the other energy domain parameters were not affected by gemination.

### 5.2. Effect of gemination in liquids
Time domain parameters for liquids were strongly correlated with gemination. Cd, V1d and Utd were in fact significantly different in singletons vs. geminates for all combinations of vowels and consonants, although the impact on Utd was typically weaker, as shown by higher p values when compared to V1d and even more to Cd.
The analysis of frequency domain parameters was carried out for liquids by analyzing both pitch F0 and formants F1, F2 and F3 in vowel frames and F0 in consonant frames. No significant variations due to gemination were observed.
Finally, in analogy with results observed for nasals, the total energy of the consonant $E_{totC}$ was the only parameter showing significant variation with gemination.

### 5.3. Comparison of acoustic correlates of gemination in nasals and liquids
Frequency and energy parameters showed a sporadic effect of gemination. For this reason, this paragraph focuses on temporal parameters only.
Table XVI summarizes mean values and standard deviations for liquids and nasals, averaged over all repetitions, speakers, consonants and vowels. Table XVI shows that consonant duration Cd is the parameter with largest relative variation across all consonant categories ($\approx$+133% in nasals, $\approx$+187% in liquids) followed by pre-consonant vowel duration V1d ($\approx$-32% in nasals, $\approx$ -41% in liquids).
Results of the analysis on the significance of time domain parameter variations for nasals (Table II) and liquids (Table IX) are in good agreement with the analysis carried out in (Esposito and Di Benedetto 1999) for stops.
Comparison in terms of Spearman Rank correlation shows that both nasals and liquids present a high negative correlation between V1d and Cd (< -0.65), while a weaker correlation is observed when the analysis is restricted to geminate words, and no correlation at all is present when only singleton words are considered.

|         |           |      | V1d    | Cd     | V2d    | Utd    | Cd/V1d |
|---------|-----------|------|--------|--------|--------|--------|--------|
| Nasals  | Singleton | Mean | 183.52 | 90.64  | 130.05 | 404.20 | 0.51   |
|         |           | StD  | 27.45  | 14.14  | 25.43  | 45.07  | 0.12   |
|         | Geminate  | Mean | 124.56 | 211.75 | 124.25 | 460.57 | 1.77   |
|         |           | StD  | 20.95  | 33.33  | 25.43  | 43.02  | 0.56   |
| Liquids | Singleton | Mean | 171.92 | 60.56  | 100.21 | 384.1  | 0.36   |
|         |           | StD  | 25.75  | 15.33  | 22.1   | 40.53  | 0.11   |
|         | Geminate  | Mean | 121.81 | 174.2  | 87.74  | 443.86 | 1.52   |
|         |           | StD  | 27.54  | 28.69  | 21.45  | 42.87  | 0.51   |

**Table XVI** - Mean values and standard deviations of the time related parameters averaged over all the repetitions, speakers, consonants and vowels for nasals and liquids.

### 5.4. Classification of geminate vs. singleton words in nasals and liquids

Results presented in Section 4 highlighted that only time domain parameters are consistently and significantly affected by gemination. Time domain parameters were thus adopted as test variables for Maximum Likelihood Classification tests (Dillon and Goldstein, 1984) of geminate vs. singleton words. Table XVII shows the classification percentage error for tests on nasals and liquids using V1d, Cd and V2d for male and female speakers and for all words combined. Results in Table XVII are in good agreement with the results of the ANOVA tests shown in Section 4; Cd, that is the parameter that presented the most significant variations with gemination also led to the lowest classification error rates. Classification tests using V1d led to higher error percentages, coherently with the weaker significance for V1d variations observed in Section 4.

|         |          | V1d  | Cd  | V2d  |
|---------|----------|------|-----|------|
| Nasals  | Combined | 10.2 | 0.5 | 41.7 |
|         | Male     | 7.4  | 0.9 | 39.8 |
|         | Female   | 16.7 | 0   | 49.1 |
| Liquids | Combined | 18.1 | 0   | 44.4 |
|         | Male     | 13.0 | 0   | 48.2 |
|         | Female   | 22.2 | 0   | 37.0 |

**Table XVII** - Percentage of singleton *vs.* geminate classification errors for nasal and liquid consonants based on unidimensional MLC tests on time domain parameters V1d, Cd and V2d for separate female and male speakers, and for all combined words.

Additional tests were carried out, to investigate the combination of multiple parameters in the classification of geminate vs. singleton words. The analysis focused on the combination of Cd and V1d. Parameters were combined in two ways. First, they were used as variables in a bidimensional MLC test, following the same approach adopted in (Esposito and Di Benedetto, 1999) for stops. Secondly, the ratio Cd/V1d was used in a unidimensional test, following what was suggested in (Pickett *et al.* 1999).
Table XVIII shows the classification error percentage for the following three cases: 1) female speakers, 2) male speakers and 3) all speakers combined.

|         |          | Bidimensional (Cd, V1d) | Unidimensional Cd/V1d |
|---------|----------|-------------------------|------------------------|
| Nasals  | Combined | 0                       | 0.5                    |
|         | Male     | 0                       | 0                      |
|         | Female   | 0                       | 1.9                    |
| Liquids | Combined | 0                       | 0.5                    |
|         | Male     | 0                       | 0.9                    |
|         | Female   | 0                       | 0.9                    |

**Table XVIII** - Percentage of singleton vs. geminate classification errors for nasal and liquid consonants in a bidimensional test using (Cd, V1d), and in an unidimensional MLC test using the Cd/V1d ratio for separate female and male speakers, and for all combined words.

Results of the bidimensional tests indicate that in nasals the introduction of V1d allowed to remove the residual classification errors observed in Table XVII when only Cd was used. In liquids the classification based on Cd was already error free, and the introduction of V1d did not affect the classification performance.

Results of the unidimensional test using the Cd/V1d ratio does not consistently lead to improved classification rates. In nasals a slight improvement was observed for male speakers when switching from C1d to C1d/V1d, while classification rate did not change for combined speakers, and actually degraded from perfect classification to a 1.9% error rate for female speakers. In liquids a small classification rate loss was observed in all groups: 0.5% for combined speakers, 0.9% for both male and female speakers.

The thresholds on Cd/V1d that led to the best classification performance in the MLC test, corresponding to the Points of Equal Probability (PEPs) between the two Gaussian distributions fitted on singleton vs. geminate data, are presented in Table XIX. Table XIX also presents the thresholds that led to the best classification performance in a heuristic test that explored all possible thresholds, without assuming Gaussian distributions for singletons vs. geminates; the heuristic test was motivated by the limited size of the set of words, that might not be properly fitted by a Gaussian distribution. Table XIX shows that the best classification performance was obtained with a threshold in the order of 0.75 in most cases for both consonant classes, the only exception being words including liquids pronounced by female speakers, for which the threshold was below 0.6.

|  |  | Cd/V1d threshold | |
| --- | --- | --- | --- |
|  |  | **MLC PEP** | **Heuristic** |
| **Nasals** | Combined | 0.8 | 0.78 |
|  | Male | 0.8 | 0.76 |
|  | Female | 0.82 | 0.78 |
| **Liquids** | Combined | 0.63 | 0.74 |
|  | Male | 0.69 | 0.74 |
|  | Female | 0.54 | 0.58 |

**Table XIX** – Thresholds for singleton *vs.* geminate classification in nasal and liquid consonants using the Cd/V1d ratio for separate female and male speakers, and for all combined words; thresholds were determined both as the Point of Equal Probability (PEP) resulting from the assumption of Gaussian distributions for the two groups of geminate and singleton words, and heuristically as the value that minimizes the number of classification errors.

These results can be compared with those presented in (Pickett *et al.* 1999) for classification of singleton vs. geminate stop consonants. Pickett (1999) found in fact by visual inspection of Cd and V1d values that classification based on Cd/V1d with an arbitrary value of 1 led to satisfactory classification error rates across different speaking rates, indicating an invariance property of Cd/V1d with speaking rate. One might thus wonder whether Cd/V1d shows a similar invariance property across different consonant categories. Results in Table XIX for nasals and liquids seem indeed to indicate that Cd/V1d may show some form of invariance across consonants, at least in terms of best classification threshold, although our threshold is lower than the one proposed in (Pickett *et al.* 1999) for stops. In order to better assess this aspect, Figure 8 presents the classification error rate in the heuristic test as a function of the Cd/V1d threshold for combined male and female speakers data for nasals and liquids.

Results in Figure 8 show that for both categories the error rate remains close to its minimum value for a wide range of Cd/V1d threshold values, due to the clear separation between singleton and separate words in terms of Cd/V1d; the value of 1 proposed in (Pickett *et al.* 1999) falls within this range for nasals but not for liquids, leaving the question of a common threshold existing for all consonant categories open. The potential role of Cd/V1d as an across-consonant classification parameter is further investigated in (Di Benedetto and De Nardis, 2020) where a similar analysis is carried out for affricates and fricatives, and the use of this parameter for classifying geminate vs. singleton consonants of all consonant classes, stops, nasals, liquids, fricatives, and affricates is tested.

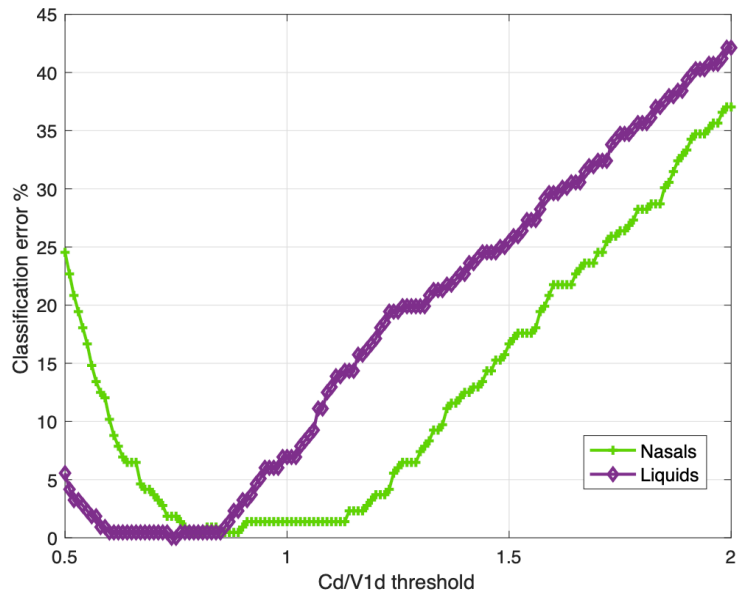

**Figure 8** - Classification error rate as a function of the Cd/V1d threshold for combined male and female speakers data for nasals vs. liquids.

## 6. Conclusions

The impact of gemination on nasal and liquid Italian consonants, based on acoustic analyses of disyllabic words (VCV vs. VCCV) in a symmetrical context of cardinal vowels [a, i, u] belonging to the GEMMA project database (GEMMA, 2019), was investigated. Time domain, frequency domain and energy domain measurements were collected in different frames within the word, corresponding to crucial events such as vowel-to-consonant transition and vowel and consonant stable portions.

The most relevant outcomes can be summarized as follows:

- a general tendency of shortening the pre-consonant vowel and of lengthening the consonant in a geminate word, that was observed in previous studies (Esposito and Di Benedetto, 1999), (Pickett et al. 1999), (Payne, 2005), (TagliaPietra and McQueen, 2010), (Turco and Braun, 2016) was confirmed for both nasals and liquids;
- a careful examination of the speech materials under study highlighted a high degree of correlation between the two aforementioned effects when considering the full set of singletons vs. geminates. A weaker correlation is already present in geminates vs. geminates, while no correlation was observed in singletons vs. singletons. This result is important since it quantifies a hypothesis suggested by Shrotiya et al., (1995), that the observed effect is related to a need of preserving rhythmical structures;
- a significant increase of pitch F0 in nasals for pre consonant vowel V1, and only for male speakers, was the only emerging effect of gemination on frequency domain parameters. No significant variation of F0 was observed for the consonant, neither in nasals nor in liquids. The result for nasals is in agreement with previous studies on gemination in nasals in other languages, in particular Pattani-Malay, where no impact of gemination on F0 was detected (Abramson, 1998), (Abramson, 1999).
- the analysis of energy-related parameters highlighted that the energy of the consonant $E_{totC}$ was significantly affected by gemination for both nasals and liquids. This result marks a clear difference with stops, for which no significant variations in energy parameters were observed (Esposito and Di Benedetto, 1999);
- the use of the primary acoustic cue Cd for classification of singletons vs. geminates led to the best classification rates for both nasals and liquids. In the case of nasals, error-free classification was obtained using Cd, while in liquids residual classification errors were eliminated by combining the primary cue with first vowel duration V1d in a bidimensional classifier;
- the Cd/V1d ratio was investigated as an across-consonant parameter for detecting gemination; satisfactory classification rates were obtained in both nasals and liquids and stops using a same threshold value. This threshold value was however different from the one proposed in previous studies for classification of gemination in stops (Pickett *et al.* 1999), questioning the invariance of Cd/V1d with consonant category. A further discussion on this aspect will be included in the companion paper by considering all five consonant categories.

# 7. Appendix – Average value and standard deviation of time domain and energy domain parameters
## 7.1. Nasals

|   |      | V1d (msecs) |       | Cd (msecs) |       | V2d (msecs) |       | Utd (msecs) |       |
|---|------|-------------|-------|------------|-------|-------------|-------|-------------|-------|
|   |      | Mean        | StD   | Mean       | StD   | Mean        | StD   | Mean        | StD   |
| a | ama  | 157.91      | 10.20 | 86.73      | 8.26  | 105.86      | 16.66 | 350.50      | 20.38 |
|   | amma | 117.77      | 15.40 | 210.08     | 27.86 | 106.27      | 21.12 | 434.13      | 41.17 |
|   | ana  | 200.98      | 20.75 | 79.57      | 12.39 | 140.46      | 21.49 | 421.00      | 37.75 |
|   | anna | 133.77      | 16.64 | 201.47     | 23.54 | 126.77      | 30.43 | 462.02      | 53.28 |
| i | imi  | 171.71      | 22.98 | 96.93      | 10.68 | 120.73      | 25.32 | 389.38      | 31.09 |
|   | immi | 118.18      | 22.75 | 227.12     | 31.99 | 120.19      | 23.58 | 465.49      | 40.57 |
|   | ini  | 187.32      | 32.61 | 87.98      | 12.85 | 141.78      | 21.53 | 417.09      | 44.05 |
|   | inni | 117.83      | 21.78 | 229.08     | 37.58 | 132.28      | 24.40 | 479.18      | 30.94 |
| u | umu  | 182.02      | 23.90 | 102.65     | 16.24 | 131.54      | 22.13 | 416.21      | 42.48 |
|   | ummu | 131.84      | 24.06 | 200.64     | 35.96 | 130.92      | 20.35 | 463.40      | 41.37 |
|   | unu  | 201.18      | 23.05 | 89.94      | 12.01 | 139.92      | 25.70 | 431.04      | 39.44 |
|   | unnu | 127.98      | 19.47 | 202.11     | 31.98 | 129.10      | 25.02 | 459.18      | 40.67 |

**Table XX** - Average and standard deviation of time domain parameters for words containing nasals in singleton vs. geminate forms, averaged over all repetitions and speakers (all values are expressed in milliseconds).

|   |      |      | $E_{totV1}$ | $P_{mV1}$ | $E_{totC}$ | $P_{mC}$ | $E_{iV1cent}$ | $E_{iV1-C}$ | $E_{iCcent}$ | $E_{iCoffset}$ |
|---|------|------|-------------|-----------|------------|----------|---------------|-------------|--------------|----------------|
| a | ama  | Mean | 102.00      | 70.06     | 89.06      | 59.61    | 94.50         | 88.33       | 83.50        | 84.28          |
|   |      | Std  | 4.51        | 4.70      | 5.23       | 5.44     | 4.75          | 5.24        | 5.58         | 5.27           |
|   | amma | Mean | 98.39       | 67.83     | 91.06      | 58.00    | 92.61         | 86.39       | 81.94        | 82.56          |
|   |      | Std  | 3.30        | 2.90      | 6.19       | 6.41     | 3.14          | 4.30        | 6.82         | 7.02           |
|   | ana  | Mean | 100.11      | 67.00     | 87.56      | 58.67    | 92.00         | 84.78       | 82.56        | 82.89          |
|   |      | Std  | 2.75        | 2.94      | 5.27       | 5.95     | 2.98          | 4.54        | 6.16         | 6.01           |
|   | anna | Mean | 98.17       | 66.72     | 92.00      | 59.11    | 92.06         | 85.11       | 82.83        | 82.28          |
|   |      | Std  | 3.13        | 2.94      | 5.47       | 5.54     | 3.23          | 4.61        | 6.09         | 5.69           |
| i | imi  | Mean | 92.72       | 60.44     | 89.11      | 59.28    | 84.72         | 84.44       | 82.89        | 82.78          |
|   |      | Std  | 5.32        | 5.30      | 7.12       | 7.07     | 5.00          | 7.29        | 6.54         | 6.70           |
|   | immi | Mean | 91.83       | 61.39     | 93.67      | 60.11    | 85.72         | 85.33       | 83.67        | 83.78          |
|   |      | Std  | 3.92        | 3.89      | 5.93       | 6.44     | 3.94          | 4.48        | 6.92         | 7.65           |
|   | ini  | Mean | 91.89       | 59.22     | 89.11      | 59.72    | 83.06         | 84.44       | 83.44        | 83.61          |
|   |      | Std  | 4.42        | 4.58      | 6.17       | 6.74     | 4.67          | 6.30        | 6.67         | 6.86           |
|   | inni | Mean | 92.11       | 61.67     | 94.11      | 60.56    | 85.61         | 84.78       | 84.28        | 84.28          |
|   |      | Std  | 4.52        | 4.09      | 5.61       | 5.91     | 4.64          | 4.45        | 6.97         | 6.93           |
| u | umu  | Mean | 94.56       | 62.11     | 87.11      | 57.11    | 87.17         | 82.11       | 80.94        | 81.94          |
|   |      | Std  | 2.84        | 2.92      | 5.75       | 6.20     | 2.71          | 5.56        | 6.62         | 6.37           |
|   | ummu | Mean | 94.67       | 63.61     | 91.06      | 58.11    | 88.89         | 82.61       | 82.00        | 82.78          |
|   |      | Std  | 3.98        | 3.79      | 4.98       | 5.52     | 4.08          | 5.14        | 5.68         | 6.99           |
|   | unu  | Mean | 94.72       | 61.67     | 88.78      | 59.28    | 86.39         | 84.44       | 83.17        | 83.33          |
|   |      | Std  | 4.29        | 4.58      | 6.41       | 6.72     | 4.55          | 6.35        | 6.72         | 7.14           |
|   | unnu | Mean | 96.22       | 65.22     | 92.33      | 59.22    | 90.56         | 86.00       | 82.61        | 83.33          |
|   |      | Std  | 2.81        | 2.37      | 4.33       | 4.81     | 2.59          | 3.02        | 4.96         | 6.23           |

**Table XXI** - Average and standard deviation of energy domain parameters for each combination of consonants [m, n], vowels [a, i, u] and singleton vs. geminate form, averaged over repetitions and speakers (values are in logarithmic form; for a list of parameters refer to Section 3.1.5).

## 7.2. Liquids

|   |      | V1d (msecs) | | Cd (msecs) | | V2d (msecs) | | Utd (msecs) | |
|---|------|------|------|------|------|------|------|------|------|
|   |      | Mean | StD | Mean | StD | Mean | StD | Mean | StD |
| a | ala  | 180.77 | 21.54 | 73.93 | 13.78 | 113.98 | 20.82 | 368.68 | 39.24 |
|   | alla | 127.25 | 24.73 | 201.69 | 28.64 | 99.59 | 16.94 | 428.53 | 34.04 |
|   | ara  | 200.54 | 29.19 | 89.99 | 19.68 | 107.99 | 28.40 | 398.53 | 37.25 |
|   | arra | 145.28 | 31.94 | 202.64 | 23.82 | 96.98 | 17.52 | 444.90 | 42.54 |
| i | ili  | 169.74 | 22.87 | 81.69 | 13.81 | 112.74 | 21.38 | 364.17 | 33.85 |
|   | illi | 127.61 | 17.51 | 204.96 | 27.17 | 112.56 | 23.97 | 445.13 | 39.21 |
|   | iri  | 191.79 | 151.89 | 95.32 | 18.22 | 115.11 | 17.03 | 402.22 | 42.16 |
|   | irri | 25.28 | 28.17 | 203.36 | 31.43 | 110.84 | 17.77 | 466.09 | 45.48 |
| u | ulu  | 176.79 | 26.04 | 90.87 | 12.62 | 120.93 | 17.85 | 388.59 | 41.56 |
|   | ullu | 123.52 | 25.83 | 207.20 | 30.54 | 103.49 | 19.68 | 434.22 | 40.67 |
|   | uru  | 188.13 | 22.26 | 85.78 | 12.89 | 108.48 | 24.91 | 382.38 | 38.68 |
|   | urru | 138.40 | 28.85 | 205.65 | 28.42 | 100.23 | 23.62 | 444.28 | 49.67 |

**Table XXII** - Average values and standard deviations (in milliseconds) of V1d, Cd, V2d and Utd for words containing liquids, averaged over all repetitions and speakers.

|   |      |      | $E_{totV1}$ | $P_{mV1}$ | $E_{totC}$ | $P_{mC}$ | $E_{iV1cent}$ | $E_{iV1-C}$ | $E_{iCcent}$ | $E_{iCoffset}$ |
|---|------|------|------|------|------|------|------|------|------|------|
| a | ala  | Mean | 106.15 | 73.85 | 94.96 | 67.72 | 98.43 | 94.21 | 91.73 | 91.62 |
|   |      | Std  | 2.29 | 1.91 | 3.80 | 4.16 | 1.71 | 3.79 | 4.24 | 4.35 |
|   | alla | Mean | 104.80 | 74.21 | 98.40 | 65.91 | 98.97 | 94.63 | 89.16 | 89.04 |
|   |      | Std  | 2.27 | 1.58 | 1.65 | 2.11 | 1.52 | 2.15 | 2.45 | 2.75 |
|   | ara  | Mean | 105.95 | 73.25 | 94.57 | 66.76 | 97.97 | 94.69 | 87.52 | 91.98 |
|   |      | Std  | 2.03 | 1.68 | 4.09 | 3.93 | 1.37 | 3.35 | 4.40 | 3.96 |
|   | arra | Mean | 104.16 | 72.93 | 94.30 | 61.86 | 97.63 | 94.21 | 83.47 | 86.48 |
|   |      | Std  | 3.19 | 2.34 | 4.29 | 4.59 | 2.37 | 2.69 | 5.25 | 4.43 |
| i | ili  | Mean | 95.82 | 63.84 | 86.48 | 58.81 | 88.56 | 83.41 | 83.14 | 82.89 |
|   |      | Std  | 3.81 | 4.00 | 4.64 | 3.73 | 3.87 | 3.42 | 4.07 | 3.68 |
|   | illi | Mean | 95.49 | 64.93 | 93.02 | 60.55 | 89.56 | 86.30 | 84.76 | 83.02 |
|   |      | Std  | 3.41 | 3.87 | 3.13 | 2.75 | 4.19 | 3.27 | 3.33 | 3.50 |
|   | iri  | Mean | 96.04 | 63.63 | 87.67 | 59.74 | 88.14 | 88.79 | 80.12 | 82.34 |
|   |      | Std  | 3.32 | 3.35 | 4.22 | 4.07 | 3.60 | 2.99 | 4.46 | 3.40 |
|   | irri | Mean | 97.20 | 66.10 | 91.06 | 58.95 | 90.25 | 92.58 | 79.56 | 82.76 |
|   |      | Std  | 3.11 | 3.67 | 3.19 | 3.55 | 4.55 | 2.29 | 5.58 | 3.64 |
| u | ulu  | Mean | 99.24 | 67.21 | 88.49 | 60.79 | 91.96 | 88.37 | 84.80 | 84.60 |
|   |      | Std  | 2.95 | 2.94 | 2.04 | 2.14 | 3.08 | 2.90 | 2.03 | 2.80 |
|   | ullu | Mean | 97.40 | 67.17 | 95.74 | 63.34 | 92.06 | 89.25 | 86.97 | 85.86 |
|   |      | Std  | 1.88 | 2.35 | 2.74 | 2.83 | 2.40 | 3.15 | 2.88 | 5.00 |
|   | uru  | Mean | 98.96 | 66.56 | 90.02 | 62.27 | 90.77 | 89.70 | 84.35 | 86.67 |
|   |      | Std  | 2.68 | 2.79 | 3.50 | 3.28 | 2.75 | 2.58 | 3.05 | 4.16 |
|   | urru | Mean | 98.66 | 67.90 | 93.88 | 61.67 | 92.30 | 92.62 | 83.84 | 84.27 |
|   |      | Std  | 3.65 | 4.08 | 3.85 | 3.70 | 4.36 | 3.50 | 4.13 | 5.87 |

**Table XXIII** - Average and standard deviation of energy domain parameters for liquids in singleton vs. geminate forms, averaged over speakers and repetitions (values are in logarithmic form).

## Acknowledgments

This work was partly funded by Sapienza University of Rome within research projects ("ex-quota 60%", "ricerca di Facoltà") in the years 1991-2020 and supported in part by the Radcliffe Institute for Advanced Study at Harvard University. The authors wish to thank Marco Mattei, Federico Macrì and Francesca Argiolas for their collaboration on the GEMMA project while they were interns at the Speech Lab at the DIET Department working toward their Master of Science degree in Electrical Engineering.